\documentclass[prx,twocolumn,english,floatfix,longbibliography]{revtex4-2}

\usepackage[utf8]{inputenc}
\usepackage{booktabs}
\usepackage{tikz}
\usepackage{graphicx}
\usepackage{amsfonts}
\usepackage{amsmath}
\usepackage{amssymb,bbold}
\usepackage{color}
\usepackage{bm}
\usepackage{bbm}
\usepackage{dsfont}
\usepackage{enumerate}
\usepackage{amsthm}
\usepackage{mathtools}
\usepackage{array}
\usepackage{mathdots}
\usepackage[colorlinks,bookmarks=false,citecolor=blue,linkcolor=blue,urlcolor=blue]{hyperref}
\usepackage{physics}
\usepackage{comment}
\usepackage{svg}
\usepackage{tikz}  
\usepackage[normalem]{ulem}

\begin{document}

\title{Error-Resilient Ground-State Preparation through Dual-Bath Cooling}

\author{Akash Malemath$^{a,b,c}$}
\author{Yannick Werner$^{a,d}$}
\author{Alexey Bochkarev$^{e}$}
\author{Sabine Müller$^{f}$}
\author{Paul Lukowicz$^{a}$}
\author{Nikolaos Palaiodimopoulos$^{a,d}$}
\author{Maximilian Kiefer-Emmanouilidis$^{a,b,c,d}$}
\affiliation{%
 $^{a}$Embedded Intelligence, German Research Center for Artificial Intelligence (DFKI), D-67663 Kaiserslautern, Germany\\
 $^{b}$Theoretical Physics, Saarland University, D-66123 Saarbr\"ucken, Germany\\
 $^{c}$Center for Quantum Technologies (QuTe), Saarland University, D-66123 Saarbr\"ucken, Germany\\
 $^{d}$ Department of Computer Science and Research Initiative QC-AI, RPTU Kaiserslautern-Landau, D-67663 Kaiserslautern, Germany\\
 $^{e}$ Department of Mathematics, RPTU Kaiserslautern-Landau, D-67663 Kaiserslautern,  Germany\\
 $^{f}$Fraunhofer Institute for Industrial Mathematics (ITWM), Kaiserslautern, Germany
}

\begin{abstract}
Error-resilient quantum algorithms (ERQA) exploit engineered dissipation to counteract errors during circuit execution, offering a complementary route toward reliable quantum computation. Here, we extend a Lindbladian ground-state preparation protocol to multi-ancilla configurations and investigate their performance for the transverse-field Ising model using ideal and noisy simulations, complemented by experiments on IQM’s Garnet superconducting processor. We introduce a generalized-normal spectral filter and analyze how filter discretization can induce unwanted upward energy transitions, an effect we refer to as aliased heating. The dual-bath configuration, with ancillas coupled to opposite boundaries of the chain, approximately doubles the convergence rate relative to the single-bath protocol and can thereby reduce the total number of two-qubit gates required to reach a target energy. In contrast, additional bulk-coupled ancillas are more susceptible to aliased heating on coarse grids and reduce the effectiveness of light-cone cancellation in the compiled circuit. Under the investigated gate-noise model, the system energy stabilizes at noise-dependent plateaus, with the dual-bath protocol tolerating approximately twice the gate-error rates of the single-bath protocol for the considered energy-accuracy threshold. While experimental performance remains constrained by hardware imperfections, our findings highlight the potential of carefully designed, minimal ancillary environments for realizing quantum algorithms stabilized by dissipation before full fault tolerance becomes available.
\end{abstract}

\maketitle

\section{Introduction}

Errors constrain the depth and accuracy of quantum computations on present-day NISQ hardware. Engineered open-system dynamics offers a complementary approach by making controlled dissipation, measurement, and resetting part of the computational protocol. Verstraete et al.~\cite{verstraete_wolf_cirac_2009} established that engineered dissipation can support universal quantum computation and entangled many-body state preparation. They further showed that, for certain classes of Hamiltonians, suitably engineered dissipative transitions can drive a system toward a target state by removing excitations generated during its evolution.

Implementing such dynamics on quantum processors to prepare the ground states of arbitrary Hamiltonians builds on
algorithms for simulating sparse Markovian evolution and Hamiltonian-based constructions of Lindbladian dynamics~\cite{childs_li_2017,cleve_wang_2017,ding_single-ancilla_2024}.
Earlier repeated-interaction models connect sequences of
system--ancilla couplings to continuous-time open-system
evolution~\cite{Attal2006}. Collision-model constructions make this connection explicit at the circuit level, enabling multipartite Markovian dynamics through elementary interactions with ancillas~\cite{cattaneo_2021}. Together, these approaches provide a framework for programmable dissipation using unitary gates and auxiliary degrees of freedom. They also motivate the development of error-resilient quantum algorithms (ERQA), whose practical usefulness depends on whether dissipative stabilization remains effective under imperfect operations.

Computational applications include the dissipative quantum eigensolver~\cite{cubitt_2023}, state preparation through engineered ancilla resetting~\cite{alcalde_puente_2024}, and dissipative search. For the latter, Allahverdyan and Petrosyan encode the sought database element as the ground state of a system coupled to a thermal bath~\cite{allahverdyan_petrosyan_2022}, while Eder et al.\ investigate search using engineered Lindbladians~\cite{eder_2025}. Alternatively, monitored search with resetting employs detection and restart strategies to modify the time to solution through non-Markovian dynamics~\cite{king_search_2026}. Related auxiliary-system approaches have also been proposed to accelerate annealing~\cite{sveistrys_2025} and suppress nonadiabatic transitions through quantum control fields~\cite{king_shortcuts_2025}. Together, these works illustrate how dissipative, measurement-induced, and coherent mechanisms can be exploited to control system dynamics, potentially improving robustness or computational performance for finite problem sizes. 

Ground-state preparation provides a direct setting for examining whether engineered relaxation remains effective under hardware noise. In a Lindbladian approach, the target state is stationary and, under suitable connectivity and spectral conditions, becomes an attractive fixed point of the dissipative dynamics~\cite{cubitt_2023,ding_simple_2026}. Ancilla-assisted implementations realize such dissipative dynamics by coupling the system to auxiliary qubits that are subsequently reset or traced out, inducing an effective nonunitary channel on the system. This mechanism connects the protocol to sympathetic cooling and related reservoir-engineering schemes~\cite{cormick_2013,moraes_neto_2017,raghunandan_2020,PhysRevLett.98.240401,PhysRevLett.115.200502}. In particular, Raghunandan et al.\ showed that a single dissipatively driven auxiliary spin can prepare low-energy states of interacting spin chains robustly against additional decoherence~\cite{raghunandan_2020}.

A central challenge in transferring these protocols to near-term quantum hardware is the competition between engineered cooling and heating introduced by the implementation itself. Although repeated dissipative steps can remove excitations, each additional layer requires imperfect quantum gates and ancilla operations that may introduce further errors. Moreover, the finite discretization of the spectral filter can induce unwanted upward energy transitions, even in the absence of hardware noise, an effect we refer to as aliased heating. Increasing the number of ancillary baths could accelerate relaxation by providing additional cooling channels, but it also increases the circuit depth and the associated hardware errors. The placement of dissipative baths can significantly influence the convergence rate~\cite{zhan_rapid_2026}, as well as the sensitivity of the cooling dynamics to filter discretization. Determining the number and placement of ancillary baths, together with the choice of spectral filter and its discretization, is therefore essential for achieving a favorable balance between cooling efficiency and implementation overhead.

In this work, we investigate these trade-offs by extending the Lindbladian ground-state preparation protocol of Ding et al.~\cite{ding_single-ancilla_2024} to multiple ancillary baths and implementing it for the one-dimensional transverse-field Ising model. We employ a generalized-normal spectral filter and examine how filter discretization and bath placement influence the balance between cooling and aliased heating. At the circuit level, we analyze how the number and placement of ancillary baths affect light-cone cancellation and derive the conditions under which accelerated convergence compensates for the additional gates required per dissipative layer. We complement this analysis with ideal and noisy simulations and experiments on IQM's Garnet superconducting processor, comparing configurations with one, two, and three ancillary baths. Our results show that, among the configurations considered, the dual-bath protocol provides the most favorable balance between cooling efficiency and implementation overhead.

The remainder of this paper is organized as follows: Sec.~\ref{Sec:theoback} introduces the theoretical framework for dissipative ground-state preparation, while Sec.~\ref{Sec:methods} describes the implementation of the protocol, including the spectral filter, circuit construction, and ancillary-bath configurations. Sec.~\ref{Sec:results} presents the numerical and experimental results, followed by a discussion of the findings and conclusions in Sec.~\ref{sec:discuss}.

\section{Theoretical Background}\label{Sec:theoback}

In this section we will briefly outline the dissipative quantum algorithm introduced in \cite{ding_single-ancilla_2024} to prepare the ground state(s) via Lindbladian approach. 
The Lindblad master equation describing the open-system dynamics is given by:
\begin{equation} \label{Eq: GKLS}
    \frac{d}{dt}\rho = \mathcal{L}[\rho] = -\mathrm{i}[H,\rho] + \sum_i \left( K_i\rho K_i^\dagger - \frac{1}{2}\{K_i^\dagger K_i, \rho\} \right),
\end{equation}
where $\rho$ is the density matrix of the system, $\mathcal{L}$ is the Liouvillian superoperator, $H$ is the system Hamiltonian, and $K_i$ denote jump operators describing dissipative transitions between energy eigenstates of $H$. Applying the second-order Trotter–Suzuki decomposition~\cite{strang1968construction,suzuki1990fractal}, the evolution of the system can be approximated as
\begin{equation} \label{Eq:trotterized_lindblad_evolution}
    \rho(t) = \mathrm{e}^{\mathcal{L}t}[\rho_0] \approx ( \mathrm{e}^{\mathcal{L}_H\tau/2}  \mathrm{e}^{\mathcal{L}_K\tau} \mathrm{e}^{\mathcal{L}_H\tau/2} )^{t/\tau} [\rho_0]
\end{equation}
where $\mathcal{L}_H$ and $\mathcal{L}_K$ denote the coherent and dissipative components of the Liouvillian, respectively, and $\tau$ is the time step. The coherent evolution is implemented through the unitary operator $\mathrm{e}^{-\mathrm{i}H\tau}$, while the dissipative evolution $\mathrm{e}^{\mathcal{L}_K \tau}$ is realized through a non-unitary quantum channel. The essence of the algorithm in \cite{ding_single-ancilla_2024} lies in engineering the latter using a single jump operator $K$, such that the resulting dynamics converge to the ground state of $H$ \cite{ding_simple_2026}. The jump operator is constructed as 
\begin{equation}\label{eq:jump}
\begin{split}
      K :&= \sum_{i,j=0}^{2^N-1} F(\lambda_i - \lambda_j)|\psi_i\rangle \langle \psi_i|A|\psi_j\rangle \langle \psi_j| \\
      &= \int_{-\infty}^{\infty} f(s)A(s)ds,
\end{split}
\end{equation}
where $\lambda_i$ and $|\psi_i\rangle$ denote the eigenvalues and eigenstates of $H$, respectively, while $A(s)=\mathrm{e}^{\mathrm{i}Hs}A\mathrm{e}^{\mathrm{-i}Hs}$ denotes the operator evolution in the Heisenberg-picture.
The coupling operator $A$ is chosen to be Hermitian such that the dissipative dynamics is ergodic \cite{ding_single-ancilla_2024, seki_dissipative_2026}. Here, ergodicity refers to the existence of a unique stationary state of the dissipative dynamics. For dissipative ground-state preparation, the coupling operator $A$ and the filter function $f$ must be chosen such that, for a given Hamiltonian $H$, this stationary state coincides with its ground state~\cite{ding_simple_2026} (see also ~\cite{PhysRevA.78.042307}). The scalar filter function $f(s)$ is defined in the time domain through the inverse Fourier transform, $f(s)=\frac{1}{2\pi}\int_{-\infty}^{\infty}F(\omega)e^{-i\omega s}\,d\omega$, where the frequency-domain filter $F(\Delta\omega)$ is real and strictly positive for $\Delta\omega=\lambda_i-\lambda_j<0$, corresponding to transitions from higher-energy states $|\psi_j\rangle$ to lower-energy states $|\psi_i\rangle$, and vanishes for $\Delta\omega\geq0$. This allows for only downward transitions in the energy landscape of the Hamiltonian, thus leading to monotonous decrease in system energy.

For implementing the action of non-unitary $\mathrm{e}^{\mathcal{L}_K\tau}$ as a quantum gate set, we highlight three central ingredients of the construction introduced in \cite{ding_single-ancilla_2024},
\begin{itemize}
    \item[1.] To avoid expensive implementation of $f(s)$, it has to be chosen such that it decays rapidly to zero as $|s| \rightarrow \infty$. This allows us to discretize and truncate the time integral over a uniform time grid using the trapezoidal rule, such that
    \begin{equation} \label{Eq:trunc_discrt_integral}
        K \approx K_s := \sum_l f(s_l) \mathrm{e}^{\mathrm{i}Hs_l}A\mathrm{e}^{\mathrm{-i}Hs_l}w_l,
      \end{equation}
    where $l= -M_s,..., M_s$ is the integer grid parameter, $s_l = l\Delta_s$ with $\Delta_s = S_s/M_s$ being grid spacing and $w_l = \Delta_s$ for $-M_s<l<M_s$ and $w_l=\Delta_s/2$ at the boundary.

    \item[2.] The discretized, generally non-Hermitian jump operator $K_s$ is embedded into a Hermitian operator $\tilde{K}_s$ acting on an augmented Hilbert space. This construction enables a unitary implementation of the dissipative channel through an ancilla-assisted dilation, consistent with the Stinespring  framework~\cite{Stinespring1955PositiveFO}:
    \begin{equation} \label{Eq:dilated_jump_op}
        \tilde{K}_s :=
        \begin{pmatrix}
            0 & K_s^\dagger \\
            K_s & 0
        \end{pmatrix}
        = \sum_{l=-M_s}^{M_s} \sigma_l \otimes A(s_l),
    \end{equation}
    where $\sigma_l = w_l\bigl(\sigma_x\mathrm{Re}(f(s_l)) 
    + \sigma_y\mathrm{Im}(f(s_l))\bigr)$, with Pauli matrices 
    $\sigma_x$ and $\sigma_y$. The ancilla is initialized in 
    the state $|0\rangle_a$, and the joint system--ancilla 
    evolution is implemented by the unitary operator
    $\mathcal{W}_s(\sqrt{\tau}) =
    \mathrm{e}^{-\mathrm{i}\tilde{K}_s\sqrt{\tau}}$.
    The reduced system state is subsequently obtained by 
    taking the partial trace over the ancilla degrees of freedom. 

    \item[3.] Following the above steps, the approximation of the 
    purely dissipative evolution after $m$ steps is given by
    \begin{equation}
    \begin{split}
      \sigma_m(\tau) &:= \mathrm{Tr}_a\Bigl(
      \mathrm{e}^{-\mathrm{i}\tilde{K}\sqrt{\tau}}
      \bigl[|0\rangle_a \langle 0|_a \\
      &\qquad \otimes \sigma_{m-1}(\tau)\bigr]
      \mathrm{e}^{\mathrm{i}\tilde{K}\sqrt{\tau}}
      \Bigr),
   \end{split}
   \end{equation}
    with $\sigma_0(\tau)=\rho_0$. Denoting the exact purely 
    dissipative evolution by $\rho_K(m\tau)=\mathrm{e}^{m\tau\mathcal{L}_K}[\rho_0]$,  the approximation error satisfies $\|\rho_K(m\tau)-\sigma_m(\tau)\|_1=\mathcal{O}(m\tau^2)$, arising from the accumulation of the local approximation  error of $\mathcal{O}(\tau^2)$ per dissipative step.
\end{itemize}

It is worth noting that, although Ref.~\cite{ding_simple_2026} proposes selecting the coupling operator $A$ from a random distribution over a set of candidate operators ${A_i}$, the optimal choice of $A$ and the filter function $f(s)$ remains problem dependent. The authors of Ref.~\cite{eder_quantum_2025} further investigates the time complexity of dissipative ground-state preparation for different choices of $A$, including operators that induce long-range and short-range transitions. 

While this construction enables dissipative ground-state preparation on gate-based quantum computers, the associated circuit depth can limit its effectiveness on current hardware. In the following section, we describe our implementation of the protocol and investigate how the choice of filter function, circuit construction, and ancillary-bath configuration affect its practical feasibility.

\section{Methods} \label{Sec:methods}

\begin{figure*}[!ht]
\centering
\includegraphics[width=1\textwidth]{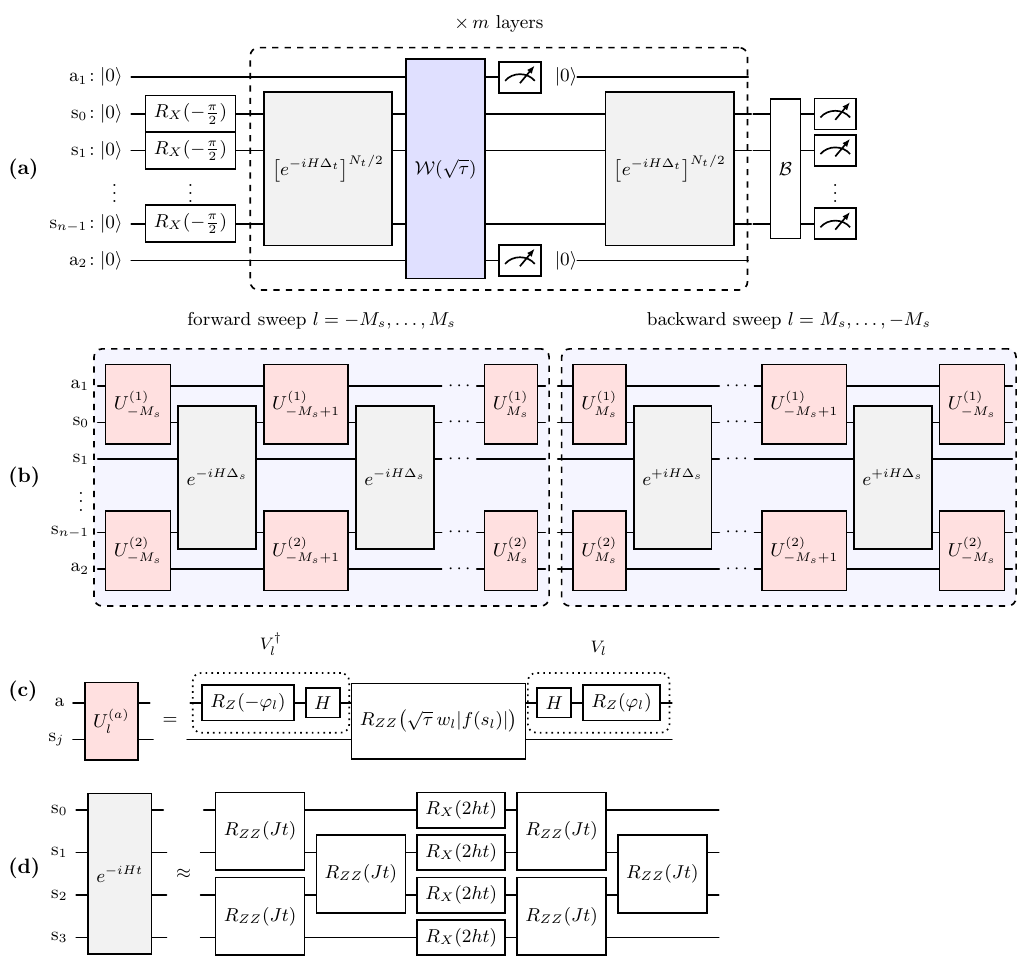}
\caption{Circuit representation of the dual ancilla dissipative ground-state preparation protocol, with one ancilla coupled to each end of the TFIM chain in Eq.~(\ref{Eq:TFIM_Hamiltonian}) (a) A single dissipative layer, repeated $m$ times, consisting of coherent evolutions, the joint system--ancilla unitary $\mathcal{W}(\sqrt{\tau})$, ancilla measurement and reset, followed by a second coherent evolution. (b) Symmetric second-order decomposition of $\mathcal{W}(\sqrt{\tau})$ into forward and backward sweeps over the discretized time grid $s_l=l\Delta_s$. At each grid point, both ancillas interact with their respective boundary sites through the operators $\mathcal{U}_l$ defined in Eq.~(\ref{eq:U_l}), with consecutive interactions separated by Hamiltonian evolution $e^{\pm iH\Delta_s}$. (c) Gate-level implementation of the elementary system--ancilla interaction. (d) Second-order Trotter decomposition of the TFIM Hamiltonian evolution, illustrated for $N=4$ qubits.}
\label{fig:full_circuit}
\end{figure*}

In this section, we adapt the dissipative ground-state preparation protocol introduced by Ding et al. \cite{ding_single-ancilla_2024} for implementation on superconducting NISQ hardware, particularly IQM's Garnet processor. We present the methodological steps aimed at reducing circuit complexity while enhancing the cooling efficiency and error resilience of the protocol. Similar to \cite{seki_dissipative_2026}, we consider the one-dimensional transverse-field Ising model (TFIM) with nearest-neighbor interactions and $N$ sites, described by the Hamiltonian
\begin{equation} \label{Eq:TFIM_Hamiltonian}
    H = J \sum_{i=0}^{N-2} Z_i Z_{i+1} + h_x \sum_{i=0}^{N-1}X_i,
\end{equation}
where $J$ denotes the exchange coupling strength and  $h_x$ the transverse magnetic field strength. 

A key design choice for the algorithm is the number and placement of ancillary baths coupled to the system. In the original scheme~\cite{ding_single-ancilla_2024}, a single ancilla couples to the system through the jump operator $K$ constructed from $A=Z_0$ at the first site of the chain.  We instead introduce two ancillas coupled to opposite ends of the chain, with $A_1=Z_0$ and $A_2=Z_{N-1}$. Since these coupling operators act on different sites, their Heisenberg-evolved counterparts commute at equal times, allowing the corresponding elementary system–ancilla interactions to be implemented independently at each grid point. As discussed further in Appendix~\ref{app:appendix_a}, the two baths scenario provide an additional relaxation channel, allowing up to two quasiparticle excitations to be removed per layer and approximately halving the number of layers required for convergence. This improvement comes at the cost of an additional system–ancilla interaction at each grid point. Introducing further ancillary baths beyond two provides only small improvements in ideal cooling performance, while resulting in worse performance on quantum hardware, as discussed in Appendix~\ref{app:appendix_a} and illustrated in Fig.~\ref{fig:2ancilla_diffloc} and Fig.~\ref{fig:3ancilla_diffloc}.

The resulting algorithm for two-bath protocol is illustrated in Fig.~\ref{fig:full_circuit}. As shown in Fig.~\ref{fig:full_circuit}(a), the system is initialized in the product state $\rho_0=\bigotimes_{i=0}^{N-1}(\mathbb{I}_i+Y_i)/2$ similar to \cite{seki_dissipative_2026}, and each dissipative layer consists of coherent evolution under $H$, followed by the joint system--ancilla unitary $\mathcal{W}(\sqrt{\tau})=e^{-i(\widetilde{K}_1+\widetilde{K}_2)\sqrt{\tau}}$, ancilla reset, and a second coherent evolution under $H$, following the Trotterized structure introduced in Eq.~(\ref{Eq:trotterized_lindblad_evolution}). Each coherent evolution is repeated $N_t/2$ times to accelerate the convergence towards the ground state \cite{seki_dissipative_2026}.
After $m$ layers, the energy expectation value is estimated from measurements of the $Z_iZ_{i+1}$ and $X_i$ observables, with the final rotation $\mathcal{B}$ implementing the corresponding measurement-basis change. Specifically, $\mathcal{B}=\mathbb{I}^{\otimes N}$ for $Z_iZ_{i+1}$ measurements and $\mathcal{B}=\mathrm{Had}^{\otimes N}$ for $X_i$ measurements, where $\mathrm{Had}$ denotes the Hadamard gate. The remaining panels illustrate the circuit decomposition required for its implementation: Fig.~\ref{fig:full_circuit}(b) shows the construction of $\mathcal{W}(\sqrt{\tau})$ through forward and backward sweeps over the discretized time grid, Fig.~\ref{fig:full_circuit}(c) presents the elementary system--ancilla interaction, and Fig.~\ref{fig:full_circuit}(d) shows the Trotterized Hamiltonian evolution. In the following subsections, we derive these circuit components and discuss their compilation into the native gate set of IQM Garnet.


\subsection{Circuit construction of dissipative block gate $\mathcal{W}(\sqrt{\tau})$}\label{sec:simpl}

The dissipative block gate with two ancilla baths $\mathcal{W}(\sqrt{\tau})$ acts jointly on the $N$ system qubits and two ancillary qubits, and is therefore represented by a $2^{(N+2)} \times 2^{(N+2)}$ unitary matrix
\begin{equation}
  \mathcal{W}(\sqrt{\tau}) = \exp\!\big[-\mathrm{i}\sqrt{\tau}\,(\tilde K_1+\tilde K_2)\big],
  \label{eq:W_two_baths}
\end{equation}
where $\tilde K_a=\sum_{l=-M_s}^{M_s}\sigma_l^{(a)}\otimes A_a(s_l)$ as seen from Eq. (\ref{Eq:dilated_jump_op}) with $a\in\{1,2\}$ labels the bath i.e., $A_1=Z_0$, $A_2=Z_{N-1}$ and $\sigma_l^{(a)}$ is the operator acting on ancilla $a$.
Following the time discretization in Eq. (\ref{Eq:trunc_discrt_integral}), we split $\mathcal{W}$ into its grid point terms with symmetric second order Trotterization corresponding to forward and backward sweep,
\begin{equation} \label{Eq:dissipative_block_split}
    \mathcal{W}(\sqrt{\tau}) \approx \prod_{l=-M_s}^{+M_s} U(s_l) \prod_{l=+M_s}^{-M_s}U(s_l) + \mathcal{O}(\Delta_s^3).
\end{equation}
Thus the total number of repetitions of $U(s_l)$ is given by $2(2M_s+1)$ and
\begin{equation}
  U(s_l)=\exp\!\Big[-\mathrm{i}\tfrac{\sqrt{\tau}}{2}\big(\sigma_l^{(1)}\otimes A_1(s_l)+\sigma_l^{(2)}\otimes A_2(s_l)\big)\Big].
\end{equation}
The two terms in $U(s_l)$ commute i.e., $A_1(s_l)$ and $A_2(s_l)$ are conjugated by the same unitary $\mathrm{e}^{\mathrm{i}Hs_l}$ and $[A_1,A_2]=0$, and $\sigma_l^{(1)}$, $\sigma_l^{(2)}$ act on different ancillas. Hence $U(s_l)$ factorises, and moving the time dependence out of the exponent gives
\begin{equation}
U(s_l)=\mathrm{e}^{\mathrm{i}Hs_l}\,\mathcal{U}_l\,\mathrm{e}^{-\mathrm{i}Hs_l};
  \qquad
  \mathcal{U}_l=U^{(1)}_l U^{(2)}_l,
  \label{eq:U_l}
\end{equation}
where 
\begin{equation}
    U^{(a)}_l=\exp\!\Big(-\mathrm{i}\tfrac{\sqrt{\tau}}{2}\,\sigma_l^{(a)}\otimes A_a\Big),
\end{equation}
and $\mathrm{e}^{\pm\mathrm{i}Hs_l}$ acts on the system only. Thus every grid point applies the two interactions $U^{(1)}_l$ and $U^{(2)}_l$ of Eq.~(\ref{Eq:dissipative_block_split}) simultaneously, on the pairs
$(\mathrm{a}_1,\mathrm{s}_0)$ and $(\mathrm{a}_2,\mathrm{s}_{N-1})$ as shown in Fig.~\ref{fig:full_circuit}.
\newline
For two neighboring grid points, the coherent Hamiltonian evolution terms in between contract to give,
\begin{equation}
  U(s_l)\,U(s_{l+1})=\mathrm{e}^{\mathrm{i}Hs_l}\,\mathcal{U}_l\,\mathrm{e}^{\mathrm{i}H\Delta_s}\,
  \mathcal{U}_{l+1}\,\mathrm{e}^{-\mathrm{i}Hs_{l+1}},
\end{equation}
since $s_{l+1} - s_l = \Delta_s$.
Applying this for Eq. (\ref{Eq:dissipative_block_split}) along both sweeps leads to similar contraction at all the grid points to give $\mathrm{e}^{\pm\mathrm{i}H\Delta_s}$, except at each end of the grid. At turning point between the sweeps $s_{M_s}$ the coherent evolutions $\mathrm{e}^{-\mathrm{i}Hs_{M_s}}$ and $\mathrm{e}^{+\mathrm{i}Hs_{M_s}}$ cancel, and the two interactions merge into $\mathcal{U}_{M_s}^2$ i.e.,
\begin{equation}
\begin{split}
  \mathcal{W}(\sqrt{\tau})\approx\,&
  \mathrm{e}^{-\mathrm{i}HS_s}\;
  \mathcal{U}_{-M_s}\,\mathrm{e}^{\mathrm{i}H\Delta_s}\,\mathcal{U}_{-M_s+1}\cdots
  \mathrm{e}^{\mathrm{i}H\Delta_s}\,\\
  &\mathcal{U}_{M_s}^{2}\,\mathrm{e}^{-\mathrm{i}H\Delta_s}\cdots
  \mathcal{U}_{-M_s+1}\,\mathrm{e}^{-\mathrm{i}H\Delta_s}\,\mathcal{U}_{-M_s}\;
  \mathrm{e}^{\mathrm{i}HS_s}.
  \label{eq:W_final}
\end{split}
\end{equation}
This is shown as the circuit in Fig. \ref{fig:full_circuit}(b), where in the forward sweep the system-bath interactions are separated by $\mathrm{e}^{-\mathrm{i}H\Delta_s}$ followed by a backward sweep separated by $\mathrm{e}^{+\mathrm{i}H\Delta_s}$. Thus $\mathcal{W}$ contains $4M_s+1$ ancilla-system interactions and $4M_s$ grid point evolutions per ancilla bath.
As discussed in \cite{seki_dissipative_2026}, the two outer evolutions $\mathrm{e}^{\mp\mathrm{i}HS_s}$ can be effectively absorbed in $\mathrm{e}^{-\mathrm{i}H\Delta_t}$ and thus can be dropped. Finally, each $\mathrm{e}^{\pm\mathrm{i}H\Delta_s}$ is implemented by a second-order Trotter step (Fig.~\ref{fig:full_circuit}(d)).
\newline
In the following subsections, we focus on designing ancilla-system interaction by
choosing a particular filter function, optimize its grid parameters for truncation and discretization of the integral in $K$s along with the time step $\tau$ in $\mathcal{W}(\sqrt{\tau})$. We examine this interaction with the aim to simplify the resulting circuit to reduce the depth for the hardware.


\subsubsection{Choice of filter function} \label{sec:filter}
Previous works have identified \emph{error function} \cite{ding_single-ancilla_2024} and \emph{Fermi-Dirac} \cite{seki_dissipative_2026} distributions as possible choices. Here, we adopt a parametric family of continuous probability distributions on the real line $\mathbb{R}$, known as the \emph{generalized normal distribution}, which adds a shape parameter to the normal distribution to control peak sharpness and tail behaviour. We choose the \emph{symmetric} version of such a distribution and it is given by
\begin{equation} \label{Eq:filter_freq_domain}
    F(\omega) = \frac{\beta}{2\alpha \Gamma(1/\beta)} \mathrm{e}^{-\left(\frac{|\omega-\mu|}{\alpha}\right)^{\beta}},
\end{equation}
where $\beta >0$ is the shape parameter, $\alpha>0$ is the scale parameter, $\mu$ determines the center of the filter function, and $\Gamma$ denotes the gamma function. The corresponding inverse Fourier transform is given by 
\begin{equation} \label{Eq:filter_time_domain}
    f(s) = \frac{\mathrm{e}^{-\mathrm{i}\mu s}}{2\pi} \frac{\beta}{\Gamma(1/\beta)} \int_{0}^{\infty} dx \cos{(\alpha s x)}\mathrm{e}^{-x^\beta}.
\end{equation}
The shape parameter $\beta$ determines both the sharpness of the filter function in the frequency domain and the decay properties of its inverse Fourier transform. For positive even-integer values of $\beta$, the filter function is smooth, resulting in a rapidly decaying $f(s)$, whereas odd-integer values lead to power-law tails. As $\beta$ increases, $F(\omega)$ approaches the ideal rectangular function of width $2\alpha$, while $f(s)$ approaches a sinc-like function with slowly decaying oscillations.
This introduces a trade-off between filter sharpness and circuit complexity. A sharper $F(\omega)$ gives a more slowly decaying $f(s)$, so a longer time window is needed to accurately approximate the jump operator. Thus, for a fixed grid spacing $\Delta_s$, this leads to an increase in the number of grid points $M_s$ and the resulting circuit depth.

\begin{figure}[!ht]
    \centering
    \includegraphics[width=\columnwidth]{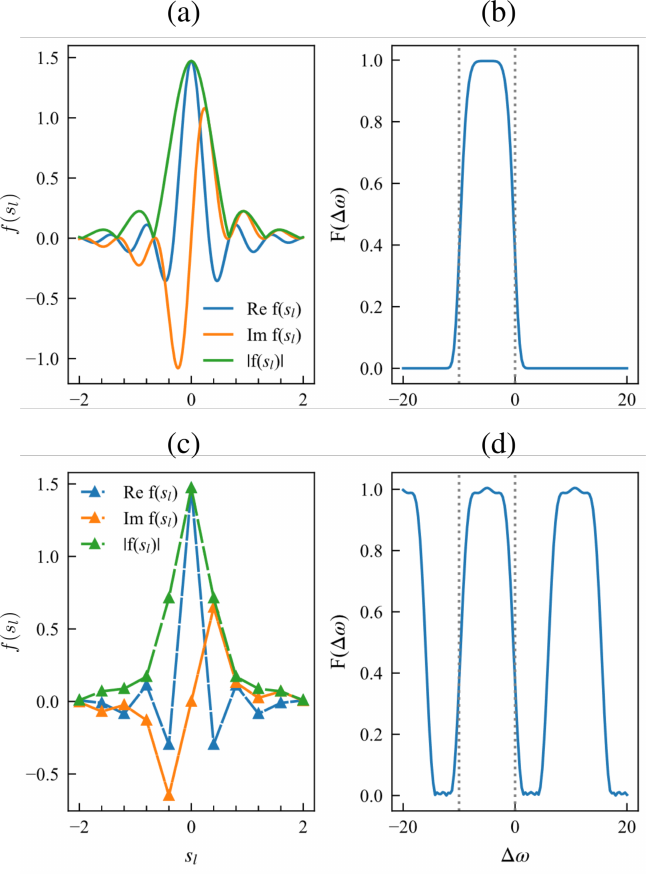}
    \caption{Generalized normal filter function with parameters $\beta=6$, $\alpha=5$, and $\mu=-5$, rescaled by a constant $C=9$. (a) Real and imaginary parts and magnitude of the inverse Fourier transform $f(s)$ from Eq.~(\ref{Eq:filter_time_domain}). (b) Corresponding frequency-domain filter $F(\omega)$ from Eq.~(\ref{Eq:filter_freq_domain}). (c) Discretized time-domain filter with grid parameters $S_s=2$ and $M_s=5$. (d) Corresponding discrete Fourier representation $F_\Delta(\omega)$, illustrating the periodic replicas of the original cooling band, including the aliased heating band at positive frequencies.}
    \label{fig:filter_function}
\end{figure}

For the implementation of the filter, $\alpha$ is chosen according to the maximum transition frequency accessible through the coupling operator $A$, while $\mu \approx -\alpha$ places the upper edge of the filter function near zero frequency, favoring downward energy transitions. However, for finite $\beta$, the filter function retains non-zero tails at positive frequencies, so upward transitions are suppressed rather than fully eliminated. The choice of $\beta$, together with the grid parameters $\{S_s,M_s\}$ in Eq.~(\ref{Eq:trunc_discrt_integral}), determines the accuracy and circuit complexity of the implementation. These parameters must therefore be selected by balancing accuracy of the desired cooling dynamics against the circuit depth supported by available NISQ hardware. In Fig.~\ref{fig:filter_function} we illustrate the chosen filter function in the time and frequency domains for $\beta=6$, $\alpha=5$, and $\mu=-5$, together with its discretized representation for $S_s=2$ and $M_s=5$.

In practice, truncating the time integral to $[-S_s,S_s]$, and discretizing the time-domain filter with grid spacing $\Delta_s$ yields a discrete Fourier representation $F_\Delta(\omega)$, which is periodic in frequency with period $2\pi/\Delta_s$. Consequently, in addition to the original cooling band $[\mu-\alpha,\mu+\alpha]$, the discretized filter contains copies of the band appearing periodically at integer multiples of $2\pi/\Delta_s$, as illustrated in Fig.~\ref{fig:filter_function}(d). The first copy of the band on the heating side $\Delta\omega>0$ starts at $2\pi/\Delta_s-(|\mu|+\alpha)$. Thus both the truncation and grid spacing effects lead to non-zero weight at positive frequencies ($\Delta\omega>0$), even in the absence of noise.
Upward transitions arising from either source lead to heating out of the ground state, an effect we refer to as \emph{aliased heating}. The filter should therefore be chosen to minimize aliased heating at a given circuit cost, not only to approximate an ideal step function.
A condition for keeping the first aliased cooling band outside the accessible positive-frequency range is
\begin{equation} \label{Eq:aliased_heating_condition}
    \Delta_s < \Delta_s^*(A) = \frac{2\pi}{|\mu|+\omega_{\max}(A) + \alpha},
\end{equation}
where $\omega_{\max}(A)$ is the largest allowed transition frequency that the coupling operator $A$ drives from the ground state.

The Fermi--Dirac window of Ref.~\cite{seki_dissipative_2026} has a time-domain kernel that decays on a time scale set by the inverse spectral gap, so suppressing the edge broadening requires a long integration window. Moreover, the aliased heating band is suppressed by choosing the grid spacing such that $\pi/\Delta_s > \|H\|$, where $\|H\|$ grows with system size $N$. Keeping the copies of this wide band away from the accessible positive-frequency range therefore requires a grid spacing that decreases with $N$. Both requirements increase the number of grid points, and hence the circuit depth.
The generalized normal filter relaxes both constraints. Its kernel $f(s)$ decays on a time scale $\sim1/\alpha$ set by the width of the cooling band, independently of the spectral gap, so a short window $S_s$ can suffice. The grid spacing can instead be chosen such that the aliased heating band lies outside the range of transition frequencies accessible through the coupling operator, as expressed in Eq.~(\ref{Eq:aliased_heating_condition}). For boundary couplings, these are single-quasiparticle transitions bounded by $2(|J|+|h_x|)$, independently of $N$ (see Appendix~\ref{app:appendix_a}).

\begin{figure}[!ht]
    \centering
    \includegraphics[width=0.45\textwidth]{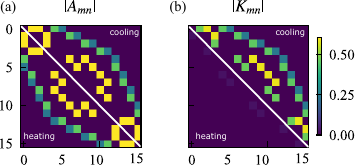}
    \caption{Matrix elements of the boundary coupling operators
in the energy eigenbasis of the TFIM Hamiltonian.
(a) Coupling operators $A_1=Z_0$ and $A_2=Z_{N-1}$
before filtering.
(b) Corresponding jump operators $K_1$ and $K_2$
after applying the filter function, illustrating the
suppression of upward energy transitions.}
    \label{fig:coupling_op_filter}
\end{figure}

The effect of the filter from Fig.~\ref{fig:filter_function} on the allowed transitions and their strengths is further illustrated in Fig.~\ref{fig:coupling_op_filter}, which compares the matrix elements of the boundary coupling operators $\{A_1,A_2\}$ with those of the corresponding jump operators $\{K_1,K_2\}$ in the energy eigenbasis of the TFIM Hamiltonian. 

\subsubsection{System--bath interaction circuit} \label{Sec:dissipative_cooling_circuit}

As seen before $\mathcal{W}$ requires an efficient circuit implementing arbitrary system-bath interaction operators of the form 
\begin{equation}\label{eq:Ul}
    U_l:=\exp\left(-i\frac{\sqrt{\tau}}{2} \sigma_l \otimes A\right).
\end{equation}
Denoting $\varphi_l:=\arg f(s_l)$, we first consider the operator
\[
    \sigma_l = w_l\begin{bmatrix}
        0 & \bar{f}(s_l)\\
        f(s_l) & 0
    \end{bmatrix} = w_l|f(s_l)|\begin{bmatrix}
    0 & e^{-i\varphi_l}\\
    e^{i\varphi_l} & 0
    \end{bmatrix}.
\]
Observe that it is a scaled 
Pauli $X$ gate in the rotated basis i.e., 
$\sigma_l = w_l|f(s_l)|R_z(\varphi_l) X R_z^\dagger(\varphi_l)$
with $R_z(\varphi):=\operatorname{diag}\Big(e^{-i\varphi/2}, e^{i\varphi/2}\Big)$.
Substituting $X={\mathrm{Had}}\cdot Z\cdot{\mathrm{Had}}$, where ${\mathrm{Had}}$ denotes the Hadamard gate, and using $e^{-i W B W^\dagger} = W e^{-iB} W^\dagger$ for arbitrary unitary $W$ and square matrix $B$ yields:
\begin{equation}\label{eq:Ul_decomposition}
    U_l = V_l \exp\left(-i\frac{\sqrt{\tau}}{2} w_l |f(s_l)|Z\otimes A\right) V_l^\dagger,
\end{equation}
where we denoted $V_l:=(R_z(\varphi_l) \cdot{\mathrm{Had}}) \otimes I$. For arbitrary coupling given by Hermitian matrix $A$ in Eq.~(\ref{eq:Ul}) is block diagonal in the basis corresponding to the eigenvectors of $\sigma_l$.
It is a quantum multiplexor \cite{shende_synthesis_2006} that applies operation $\exp(\mp i(\sqrt{\tau}/2)w_l|f(s_l)|\, A)$ to the main register depending on which eigenstate of
$\sigma_l$ the ancilla is in. In general, the exponential in Eq. \eqref{eq:Ul_decomposition} is an operator of the form
\[
    M=\ket{0}\bra{0}\otimes e^{-i\theta A} + \ket{1}\bra{1}\otimes e^{+i\theta A}
\]
with $\theta:=\frac{\sqrt{\tau}}{2} w_l |f(s_l)|$. It can be constructed from operator $e^{-i\theta A}$ and a unitary $R$
that would satisfy $RAR^\dagger=-A$. Indeed, since $Re^{-i\theta A}R^\dagger = e^{-i\theta RAR^\dagger} = e^{+i\theta A}$, and with $CR$ denoting $R$ controlled 
on the ancilla, the circuit
\[
    M = CR\, \left(I \otimes e^{-i\theta A}\right)\, CR^\dagger
\]
implements precisely the desired quantum multiplexor. For the case $A=Z_i$ at position $i$, this yields the circuit
\begin{equation} \label{Eq:U_l}
\begin{split}
    U_l &= V_l \, CX_{q_i|a}\, e^{-i(\sqrt{\tau}/2)w_l|f(s_l)|Z_i}\, CX_{q_i|a}\, V_l^\dagger\\
    &= V_l \, R_{zz}(\sqrt{\tau}w_l|f(s_l)|) V_l^\dagger,
\end{split}
\end{equation}
where we chose $R=X_i=X_i^\dagger$ and denote $CX_{q_i|a}$ as CNOT gate acting on
the $i$-th system qubit controlled by the ancilla $a$. In practice, we used
the gate 
$R_{zz}(\xi):=\exp(-i\frac{\xi}{2}Z\otimes Z_i)$. The circuit representation of the same is shown in Fig. \ref{fig:full_circuit}(c).


\subsection{Optimizing dissipative timestep}

Having established the circuit implementation of the dissipative block, we now examine how the choice of the dissipative timestep $\tau$ influences the resulting quantum channel. For the following derivation, we consider a single ancilla coupled to the system through a jump operator $K$, assumed to generate ergodic dissipative dynamics. Using the Kraus formalism, we derive the transition probabilities associated with the ancilla measurement outcomes and discuss how $\tau$ can be chosen to maximize the jump probability. As shown in Ref.~\cite{seki_dissipative_2026, ding_simple_2026}, the ground state is a fixed point of the resulting quantum channel $\Lambda_K$,
\begin{equation}
    \Lambda_K [|\psi_{gs}\rangle \langle \psi_{gs}|] = |\psi_{gs}\rangle \langle \psi_{gs}|,
\end{equation}
where the subscript $K$ corresponds to the jump operator.
The unitary operation $\mathcal{W}(\sqrt{\tau}) = \mathrm{e}^{-\mathrm{i}\tilde{K}\sqrt{\tau}}$ gives rise to following Kraus operators
\begin{equation}
    M_0 = \langle0_a|\mathcal{W}(\sqrt{\tau}) |0_a\rangle\,; \qquad M_1 = \langle1_a|\mathcal{W}(\sqrt{\tau}) |0_a\rangle,
\end{equation}
such that the action of the quantum channel at the m-th step is given by $\Lambda_K[\rho_{m-1}] = M_0\rho_{m-1}M^\dagger_0 + M_1\rho_{m-1}M^\dagger_1$. Further simplification leads to,
\begin{equation}
\begin{split}
M_0 &= \sum_n \frac{(-1)^n \tau^n}{(2n)!}
(K^\dagger K)^n
= \cos{\left(\sqrt{\tau K^\dagger K}\right)}, \\
M_1 &= -\mathrm{i}\sqrt{\tau}\sum_n
\frac{(-1)^n\tau^n K}{(2n+1)!}(K^\dagger K)^n \\
&= -\mathrm{i}\sqrt{\tau}K
\operatorname{sinc}{\left(\sqrt{\tau K^\dagger K}\right)}.
\end{split}
\end{equation}

By singular value decomposition of jump operator $K = \sum_{j} \kappa_j|u_j\rangle \langle v_j|$ such that $K^\dagger K = \sum_j \kappa_j^2 |v_j\rangle \langle v_j|$, thus we have
\begin{equation}
     M_0 = \sum_j \cos{\theta_j}|v_j\rangle\langle v_j|\,; \quad M_1 = -\mathrm{i}\sum_j \sin{\theta_j}|u_j\rangle\langle v_j|,
\end{equation}
with $\theta_j = \sqrt{\tau}\kappa_j$. The probability of dissipation at each timestep is given by,
\begin{equation}
    P_\text{d} = \mathrm{Tr}[M_1 \rho M_1^\dagger] = \sum_j \sin^2{\theta_j} \langle v_j|\rho|v_j\rangle.
\end{equation}
The probability $P_\text{d}$ can be maximized by choosing the rotation to be $\theta_j = \pi/2$. For an arbitrary choice of Hermitian coupling operator $A$, the dissipative time step $\tau$ can be chosen independently of the Hamiltonian evolution time step as
\begin{equation} \label{Eq:max_dissipation_tau}
    \tau = \left(\frac{\pi}{2\,\text{max}(\kappa_j)}\right)^2.
\end{equation}

Having established the timestep criterion for a single
jump operator, we now consider how the coupling location
affects its applicability. For the TFIM considered here,
boundary and bulk jump operators exhibit different
singular-value structures. As shown in Appendix~\ref{app:appendix_a},
under an ideal downward-only filter, all non-zero
singular values of each boundary jump operator are
identical. Consequently, the timestep in
Eq.~\eqref{Eq:max_dissipation_tau} simultaneously
maximizes the jump probability for all non-zero
singular modes. In contrast, bulk jump operators
generally exhibit several distinct non-zero singular
values, so a single dissipative timestep cannot, in
general, simultaneously optimize all the corresponding
transition channels.

As further shown in Appendix~\ref{app:appendix_a}, the finite single-particle
energy band of the TFIM and the choice of filter function
provide system-size-independent bounds on the largest
singular value of the boundary jump operators and,
consequently, on $\tau_{\max}$. Numerically, we observe
only a weak dependence of $\tau_{\max}$ on $N$,
approaching a value of approximately $2.98$ for the
discretized filter with $S_s=2$ and $M_s=5$.
Moreover, reflection symmetry ensures that the same
timestep applies to both boundaries. In practice,
$\tau_{\max}$ can therefore be estimated from a
sufficiently small system or through a $2N\times2N$
free-fermion calculation, avoiding the diagonalization
of the full many-body Hamiltonian.

Although the above derivation considers a single ergodic jump operator, the multi-bath protocol does not require each jump operator to individually enable ergodicity. Instead, the combined dissipative dynamics generated by the set of jump operators must be ergodic. Different baths may therefore provide complementary relaxation channels, even when their individual dissipative dynamics admit additional stationary states.


\section{Results} \label{Sec:results}

We evaluate the dissipative ground-state preparation algorithm for the one-dimensional transverse-field Ising model defined in Eq.~\eqref{Eq:TFIM_Hamiltonian}, with exchange coupling $J=-1$ and transverse-field strength $h_x=-1.2$. We first investigate convergence toward the ground state under ideal noiseless conditions, along with circuit complexity in terms of gate counts and the number of dissipative layers. We then analyze the robustness of the algorithm against gate noise and its performance on superconducting quantum hardware.

Unless stated otherwise, we use the generalized-normal filter function with parameters $\beta=6$, $\alpha=5$, and $\mu=-5$ and the grid parameters are set to $S_s=2$ and $M_s=5$, respectively as shown in Fig.~\ref{fig:filter_function}. For hardware experiments, we instead use $S_s=0.8$ and $M_s=2$ to reduce circuit complexity and mitigate accumulated gate errors. Both choices correspond to the same grid spacing $\Delta_s=0.4$. In the noisy simulations, we investigate configurations with multiple ancillary baths coupled to the system through local operators $A_a=Z_{j_a}$ at selected sites $j_a$. Following Eq.~\eqref{Eq:max_dissipation_tau}, the dissipative timestep is set to $\tau=2.97$, while the Trotterized Hamiltonian evolution uses an independently chosen timestep $\Delta_t=0.25$. The same filter, grid parameters, as well as the dissipative timestep, are used for all ancillary baths within each configuration.


\subsection{Aer simulations} \label{Sec: Aer Simulations}

We first evaluate the performance of the algorithm using Qiskit Aer for system sizes $N=4,6,8$ qubits and $10^4$ measurement shots per circuit for the single- and dual-bath configurations.

\begin{figure}
    \centering
    \includegraphics[width=\columnwidth]{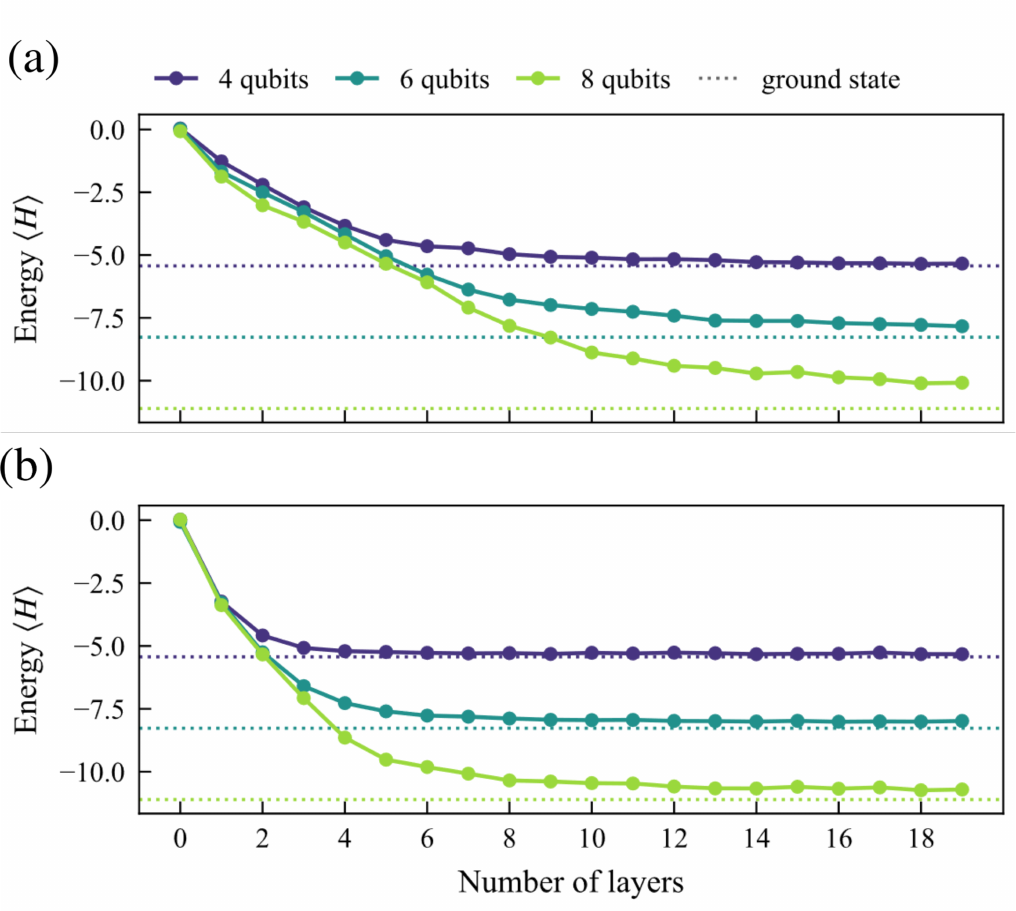}
    \caption{Simulation results for the transverse-field Ising model ($J=-1$, $h_x=-1.2$) with system sizes $N=4,6,8$ qubits, obtained using Qiskit Aer with $10^4$ measurement shots per circuit. (a) Energy convergence for the single-ancilla configuration, with the bath coupled to the first system qubit $(a_1,s_0)$. (b) Energy convergence for the two-ancilla configuration, with baths coupled to opposite ends of the chain $(a_1,s_0)$ and $(a_2,s_{N-1})$. The horizontal dashed lines indicate the exact ground-state energies for the corresponding system sizes.}
    \label{fig:2anc_1anc_aer_energy_curves}
\end{figure}

\begin{figure}
    \centering
    \includegraphics[width=\columnwidth]{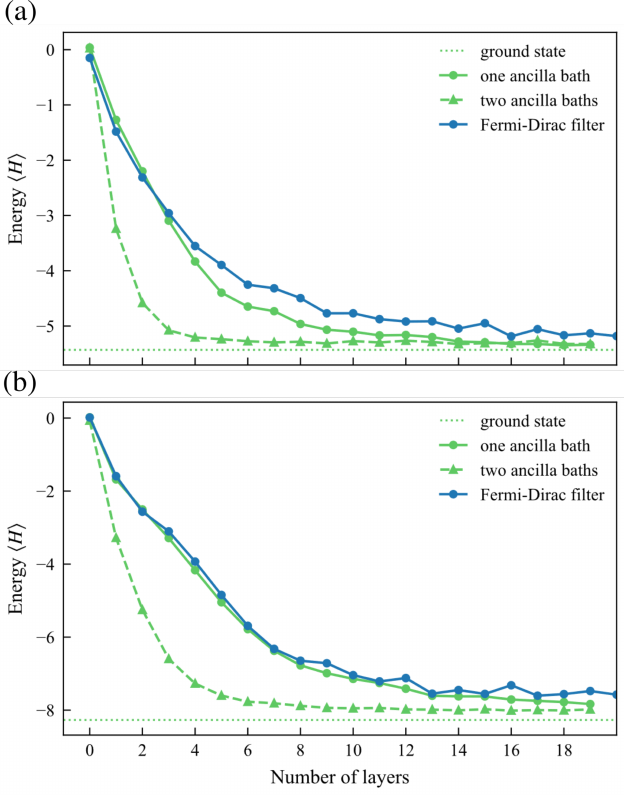}
    \caption{Comparison of the energy convergence of our single- and two-ancilla protocols with the simulation results obtained using the Fermi--Dirac filter reported by Seki et al.~\cite{seki_dissipative_2026} for (a) $N=4$ and (b) $N=6$ qubits. For the single-ancilla configurations, the protocols differ only in the choice of filter function and its associated parameters, while all other parameters are kept identical. The horizontal dotted lines indicate the exact ground-state energies for the corresponding system sizes.}
    \label{fig:aer_comparision_quantinuum}
\end{figure}

Figure~\ref{fig:2anc_1anc_aer_energy_curves} shows the convergence of the system energy toward the ground-state energy as a function of the number of dissipative layers for both configurations. As discussed earlier in Sec.~\ref{Sec:methods}, the introduction of the second ancilla accelerates convergence, approximately halving the number of layers required to reach a given energy accuracy. We further compare the performance of our protocol with the Aer simulation results obtained using a Fermi--Dirac filter in Ref.~\cite{seki_dissipative_2026}, as shown in Fig.~\ref{fig:aer_comparision_quantinuum}. While the single-ancilla configuration exhibits comparable convergence behavior, the two-ancilla scheme demonstrates a clear improvement over the reference implementation, achieving similar energies with substantially fewer dissipative layers.

\begin{figure}
    \centering
    \includegraphics{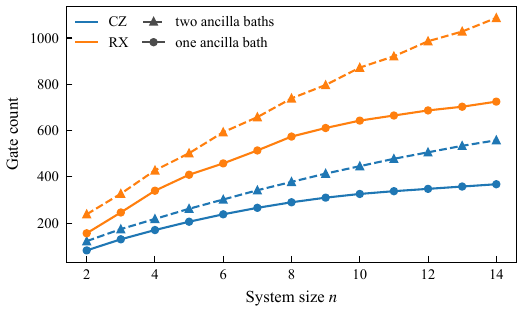}
\caption{Gate count scaling shown for a single layer with increasing system size $N$. The plot is shown after transpilation to IQM's native two-qubit and single qubit gate set, CZ and PRX (RX). The solid curves correponds to the case of single ancilla bath and the dashed curves to the dual ancilla baths. The counting analysis is performed for the generalized normal filter function with the same grid parameters as shown in Fig.~\ref{fig:filter_function}.}
    \label{fig:2anc_1anc_gate_counts_scaling}
\end{figure}

We next examine the circuit complexity of the algorithm in terms of the number of native gates per dissipative layer. Figure~\ref{fig:2anc_1anc_gate_counts_scaling} shows the gate count after the circuit construction described in Sec.~\ref{sec:simpl} and compilation to IQM Garnet's native gate set, consisting of controlled-Z (CZ) and phased rotation-X (PRX) gates. The transpilation is performed via Qiskit's \emph{`transpile\_to\_IQM()}'~\cite{qiskit2024} function. 

\begin{figure*}[!t]
\centering
\includegraphics[width=1\textwidth]{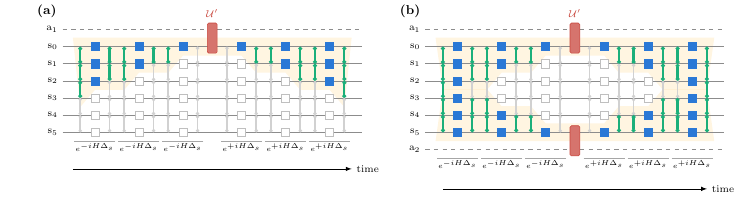}
\caption{Light cone cancellation (LCC) around the turning point of the sweep in $\mathcal{W}(\sqrt{\tau})$, shown for $N=6$ and the three Trotter steps on either side of the merged interaction $\mathcal{U}'=\mathcal{U}_{M_s}^2$ (red) (see Eq.~(\ref{eq:W_final})). Each Trotter step consists of a half layer of $R_{ZZ}$ gates (bars), a layer of $R_X$ gates (squares) and another $R_{ZZ}$ half layer (see Fig. \ref{fig:full_circuit}(d)). Gates on the left and right at equal distance from $\mathcal{U}'$ are mutual inverses. After compilation, coloured gates remain and grey gates cancel. The remaining gates fill the light cone of the coupled site (shaded), which grows by one site per Trotter step. (a)~One ancilla coupled to $\mathrm{s}_0$. (b)~Two ancillas coupled to $\mathrm{s}_0$ and $\mathrm{s}_{N-1}$ where the light cones grow from both ends and fewer gates cancel.}
\label{fig:lightcone}
\end{figure*}

For a single ancillary bath, the CZ gate count increases sublinearly over the investigated system sizes. This behavior can be explained by light-cone cancellation (LCC), a technique also used in out-of-time-order correlator circuits~\cite{doi:10.1126/science.abg5029} and light-cone renormalization-group methods~\cite{Enss_2012,Kiefer_2017} (see also Ref.~\cite{lee_light_2026}). In the dissipative block $\mathcal{W}(\sqrt{\tau})$, as seen in Eq.~\eqref{eq:W_final}, after Trotterization, each $e^{\pm iH\Delta_s}$ is decomposed into nearest-neighbor gates, enabling cancellations between the forward and backward sweeps. Around the turning point, the merged interaction $\mathcal{U}_{M_s}^2$ is enclosed by forward and inverse Hamiltonian evolutions. The region influenced by this interaction under successive Hamiltonian evolution steps defines its light cone. Gates outside this region commute through the intermediate operations and cancel with their inverses, as illustrated in Fig.~\ref{fig:lightcone}(a). For a fixed number of Trotter steps, the light cone extends over a finite number of sites, so the gate-count contribution from the dissipative block saturates with increasing system size. Beyond the light-cone extent, the total CZ gate count increases linearly due to the intermediate Hamiltonian evolutions, as derived in Appendix~\ref{app:appendix_c}.

Introducing a second ancilla adds $4M_s+1$ ancilla--system interactions per dissipative layer. Moreover, because the two ancillas couple to opposite ends of the chain, their light cones propagate inward, reducing the number of gates that can be canceled, as illustrated in Fig.~\ref{fig:lightcone}(b). This explains the higher CZ gate counts per layer observed for the dual-bath configuration in Fig.~\ref{fig:2anc_1anc_gate_counts_scaling}. As shown analytically in Appendix~\ref{app:appendix_c}, once the two light cones become spatially separated, the additional gate count introduced by the second bath becomes independent of the system size. Consequently, the relative gate-count overhead decreases with increasing $N$. More generally, the effectiveness of LCC depends on the number and placement of ancillary baths. In particular, coupling additional ancillas to bulk sites can substantially reduce the effectiveness of LCC, leading to increased circuit depth.

The increased gate count per dissipative layer does not necessarily translate into a higher total gate count for ground-state preparation. In Appendix~\ref{app:appendix_c}, we derive the convergence speed-up required for the dual-bath protocol to compensate for the additional CZ gates per layer. The approximately twofold acceleration observed in Fig.~\ref{fig:2anc_1anc_aer_energy_curves} exceeds this threshold within the considered gate-count model. For example, the $N=5$ system exhibits a $52\%$ reduction of the two-qubit gate-count. The dual-bath protocol thus achieves faster convergence at a lower overall gate cost, despite its increased circuit complexity per layer.


\definecolor{rxcol}{RGB}{42,120,214}
\definecolor{zzcol}{RGB}{27,175,122}
\definecolor{Uc}{RGB}{200,70,60}
\definecolor{conecol}{RGB}{255,236,200}


\subsection{Noisy simulations and IQM hardware} \label{sec:noisy-sim}
We now assess the error resilience of the algorithm through noisy simulations at varying noise levels. We construct a noise model based on the reference median error rates reported for IQM's Emerald processor by scaling two-qubit CZ and single-qubit PRX error rates by a common factor $\eta$, such that $\epsilon_{\mathrm{CZ}}=\eta \times\epsilon_{\mathrm{CZ}}^{(0)}$ and $\epsilon_{\mathrm{PRX}}=\eta\times\epsilon_{\mathrm{PRX}}^{(0)}$, where $\epsilon_{\mathrm{CZ}}^{(0)}$ and $\epsilon_{\mathrm{PRX}}^{(0)}$ denote the reference error rates. This preserves their ratio $r_\textrm{err}=\epsilon_{\mathrm{CZ}}/\epsilon_{\mathrm{PRX}}\approx 7$ obtained from the calibration data on the day of the experiment. By varying $\eta$, we examine how the attainable energy depends on hardware noise and compare the error resilience of configurations with different numbers of ancillary baths. This analysis also allows us to estimate the reduction in gate-error rates required to obtain a threshold accuracy relative to the noiseless implementation. Additionally, we include a fixed readout error of $2\%$ as seen on the day of the experiment, based on IQM's calibration data, to account for measurement imperfections. Additionally, noiseless density-matrix simulations are used as reference to benchmark the effects of hardware noise.

\begin{figure*}[!ht]
    \centering
    \includegraphics{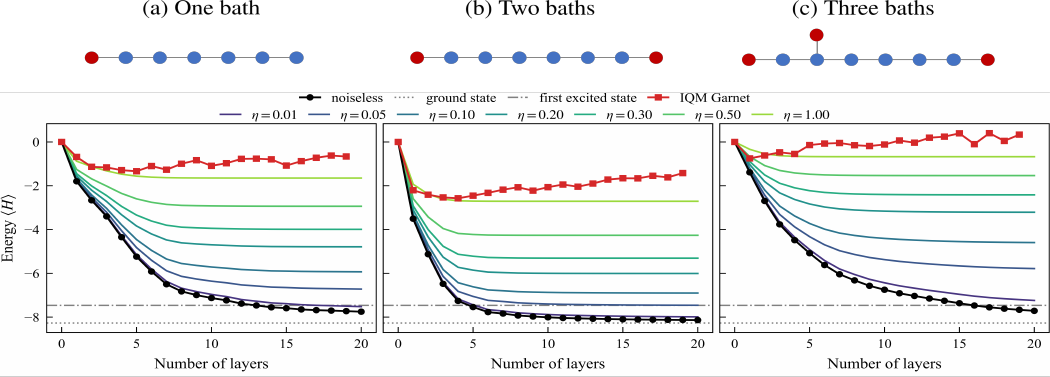}
    \caption{Noisy simulation results for a system of $N=6$ qubits at different noise scales $\eta$. The energy expectation value $\langle H\rangle$ is shown as a function of the number of dissipative layers $m$ for (a) the single-bath configuration, (b) the dual-bath configuration with ancillas coupled to opposite ends of the chain, and (c) the three-bath configuration with an additional ancilla coupled to the second system qubit. The schematic diagrams above each panel illustrate the corresponding ancilla placements. The red curves represent experimental results obtained on IQM Garnet, while the noiseless curves are obtained using density-matrix simulations.}
    \label{fig:noisy_curves}
\end{figure*}

Figure~\ref{fig:noisy_curves} shows the energy as a function of the number of dissipative layers for the model described above, for $N=6$ qubits with (a) single-bath, (b) dual-bath, and (c) three-bath configurations. We observe that the gate noise leads to energy approaching a noise dependent plateau, such that lower noise levels (i.e. smaller $\eta$) lead to plateaus closer to the ground state. Notably, the energy does not rise again with additional dissipative layers, suggesting that the engineered cooling counteracts noise-induced heating and stabilizes the system at a finite energy.
The dual-bath configuration in Fig.~\ref{fig:noisy_curves}(b) reaches lower energy plateaus than the single-bath configuration in Fig.~\ref{fig:noisy_curves}(a) at comparable noise levels, despite the increased gate count per layer. In contrast, the three-bath configuration in Fig.~\ref{fig:noisy_curves}(c) results in higher energy plateaus, indicating that the additional circuit complexity outweighs the benefits of additional cooling under the considered noise model. Further analysis of this multi-ancillary bath configuration is provided in Appendix~\ref{app:appendix_b}. Comparing the plateau energies with the ground-state and first-excited-state energies (see the horizontal dashed lines in Fig.~\ref{fig:noisy_curves}) further highlights the advantage of the dual-bath configuration, which maintains energies below the first-excited-state level over a wider range of noise levels.

The poor performance of the three-bath configuration can be partly attributed to the different fermionic structure of boundary and bulk couplings. Unlike boundary jump operators, which generate single-quasiparticle transitions under an ideal downward-only filter, bulk jump operators can induce multi-quasiparticle transitions that are more susceptible to aliased heating on coarse discretization grids, as discussed in Sec.~\ref{sec:filter}. Indeed, a finer grid does improve the convergence of the three-bath configuration, as demonstrated in Appendix~\ref{app:appendix_b}.

Figure~\ref{fig:noisy_curves} also includes experimental results obtained on IQM Garnet for the different ancillary-bath configurations. Unlike the simulated curves, which approach stable energy plateaus, the hardware results exhibit a modest initial decrease followed by an increase in energy with additional dissipative layers. As expected, during the initial layers, the hardware results approximately follow the simulated curves at noise scale $\eta=1$, corresponding to the reference IQM Emerald error rates. This initial agreement is consistent with the comparable error ratio of $r_\textrm{err}\approx9$ for IQM Garnet on the day of experiment. However, the subsequent increase in energy observed on IQM Garnet suggests that the simplified noise model does not fully capture the errors affecting the implemented dynamics. Possible contributing factors include thermal relaxation, decoherence during idle periods, and imperfect ancilla resets.

\begin{figure}
    \centering
    \includegraphics[width=\columnwidth]{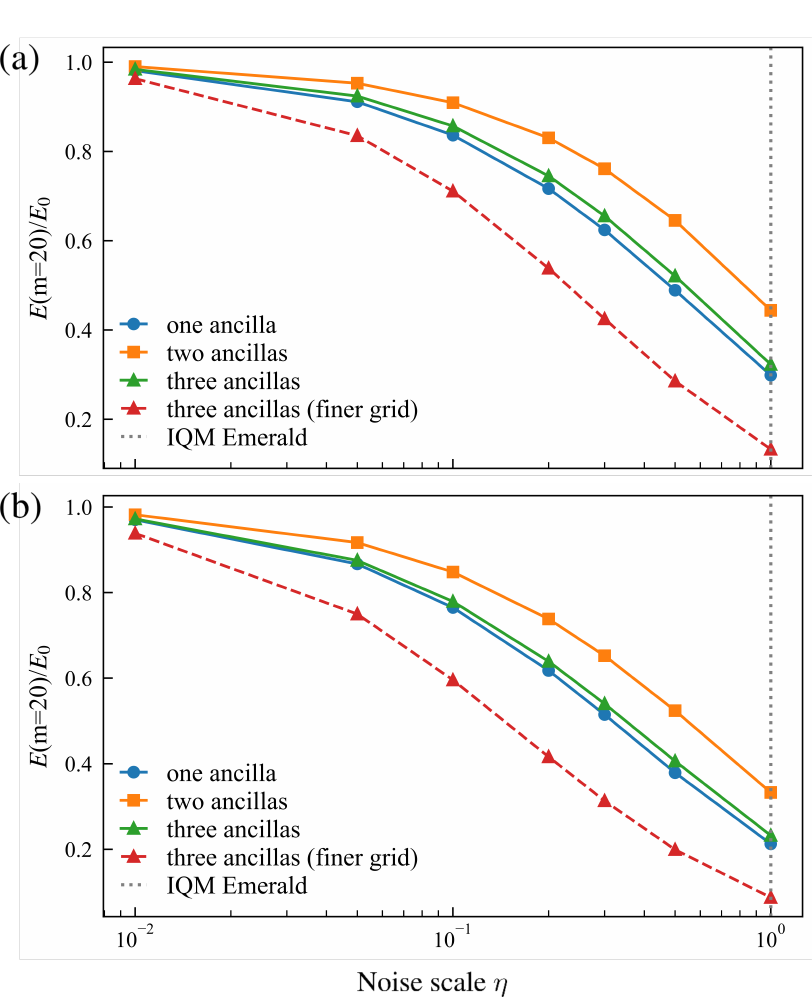}
    \caption{The plots showcase the ratio of energy obtained in presence of noise at $m=20$ layers i.e., the plateau height in Fig. \ref{fig:noisy_curves} with respect to the noiseless curves, as a function of noise scale $\eta$. The top plot corresponds to the system size of $N=4$ and bottom plot to $N=6$ qubits. The dotted curves here correspond to the finer grid scenario i.e., $M_s=10$ for the three-bath configuration. The vertical grey line represents the current best IQM hardware quality.}
    \label{fig:noise_scaling}
\end{figure}

Figure~\ref{fig:noise_scaling} quantifies the dependence of the energy on noise level through the ratio $E(m=20)/E_0$, where $E(m=20)$ is the energy after 20 dissipative layers at noise scale $\eta$, and $E_0$ is the corresponding noiseless energy. For a maximum relative deviation of $20\%$ from the noiseless energy as a threshold, we find that the single-bath protocol requires gate-error rates 20 times lower than the reference IQM Emerald values, whereas the dual-ancilla protocol requires a reduction by a factor of only 10. At this threshold, the dual-bath protocol therefore tolerates gate errors roughly twice as large as the single-bath protocol. Furthermore as expected, we observe that the three-bath configuration performs worse than the dual-bath configuration on the same coarse grid due to aliased heating. Although finer discretization ($M_s=10$) suppresses aliased heating, the associated increase in gate count leads to even worse accuracies. Overall, among the single-, dual-, and three-bath configurations investigated, the dual-bath protocol achieves the best balance between cooling performance and noise resilience.

\section{Discussion and Outlook} \label{sec:discuss}

We have adapted the single-ancilla Lindbladian cooling protocol of Ding et al.~\cite{ding_single-ancilla_2024} to multi-ancilla configurations and implemented it for the transverse-field Ising model on superconducting NISQ hardware. This work highlights that the spectral filter, its discretization, and the number and placement of the ancillary baths together determine the cooling efficiency and circuit complexity of the algorithm. The practical hardware implementation thus calls for joint optimization of these algorithmic parts.

The limited circuit depth accessible on current hardware restricts the implementation of spectral filters to coarse time grids. On such grids, the discretized filter acquires spurious weight at positive frequencies, through edge broadening from the finite integration window and periodic band copies from the finite grid spacing. The resulting jump operator then excites the ground state even in the absence of hardware noise. This aliased heating is therefore an intrinsic error source of any dissipative protocol based on a discretized spectral filter. It suggests a different design principle that unwanted transitions need only be suppressed within the frequency range accessible to the choosen coupling operators, rather than across the full many-body spectrum of the Hamiltonian. The generalized-normal filter introduced here is designed to follow these criterion i.e., its time-domain kernel decays on a scale set by the width of the cooling band rather than by the spectral gap, so short integration windows and coarse grids suffice.
The accessible frequencies are dependent on the location of baths coupled, and this links the filter directly to the bath configuration. As shown in Appendix~\ref{app:appendix_a}, a boundary coupling generates only single-quasiparticle transitions, whose frequencies are bounded independently of the system size. A bulk coupling, in contrast, also drives multi-quasiparticle transitions that extend towards the many-body bandwidth, making it far more susceptible to aliased heating. For boundary baths, the same grid parameters can therefore be retained as N grows. Together with the bounded, system-size-independent optimal dissipative timestep derived in Sec.~\ref{Sec:methods}\,B, this allows the protocol parameters to be determined on small systems and transferred to larger ones. Bulk baths can still accelerate cooling once the grid is fine enough, but the required depth limits their usefulness on current hardware. Bath placement also affects the compiled circuit. Each additional bath, and in particular each bulk bath, opens a further light cone and reduces the number of Hamiltonian-evolution gates removed by light-cone cancellation. The best configuration is therefore not the one that converges fastest under ideal conditions, but the one that cools most effectively within the available circuit depth.

Among the configurations investigated, coupling one bath locally to each boundary of the chain with coupling operator $Z$ meets this criterion best. The dual-bath protocol requires fewer than half as many layers as the single-bath protocol to reach a given energy, and it achieves lower energies than all other multi-bath configurations, with and without hardware noise. As shown in Appendix~\ref{app:appendix_c}, this faster convergence outweighs the higher gate count per layer, so the total number of two-qubit gates needed to reach a target energy is reduced, for example for system of $N=5$, the two-qubit gate count reduction is $52\%$. Moreover, once the light cones of the two baths become spatially separated, the gate overhead of the second bath becomes independent of the system size and its relative weight decreases as N grows.

The above discussion between cooling performance and gate count determines the behaviour of the protocol in the presence of noise. Under the investigated noise model with gate noise, the energy approaches a noise-dependent plateau, consistent with a balance between engineered cooling and noise-induced heating. The improved noise resilience of the dual-bath configuration, despite its higher gate count per layer, demonstrates that the additional cooling can more than compensate for additional errors introduced. For the parameters considered, requiring an threshold energy within 20\% of the noiseless result after m = 20 layers, the single-bath protocol needs gate-error rates about 20 times below the IQM Emerald reference values, whereas the dual-bath protocol needs only a ten-fold reduction. The dual-bath configuration therefore tolerates approximately twice the gate-error rates of the single-bath configuration.

The experiments on IQM Garnet show that current hardware poses challenges beyond those captured by this noise model. Instead of energy stabilizing on a plateau, the measured energy rises again after an initial decrease. A depolarizing model cannot produce this behaviour, which points to additional error mechanisms that accumulate over repeated layers such as amplitude damping during idle periods, imperfect ancilla resets, coherent over-rotations, and calibration drift.

Circuit depth remains a major bottleneck for extending such dissipative protocols to larger systems on NISQ hardware. Although the light-cone cancellations already remove a substantial fraction of the Hamiltonian evolution gates within the dissipative block, its repeated structure makes it a natural target for further improvements through variational or approximate compilation of local time-evolution circuits~\cite{zhang2026scalable, robertson2025aqctensor}. Such compression techniques are useful, provided that the resulting approximation errors do not significantly alter the stationary state of the implemented channel. Establishing this trade-off requires studying how coherent compilation errors affect both the stationary state and the convergence properties of the dissipative dynamics. Beyond the TFIM, a natural direction is to identify, for broader classes of Hamiltonians, coupling operators that cool efficiently, multi-ancilla configurations that access only a narrow range of transition frequencies, and preserve light-cone cancellation.

More broadly, our results show that the noise resilience of dissipative quantum algorithms depends not only on the strength of the engineered dissipation but on how its relaxation channels are constructed. By exploiting the spectral and spatial structure of the model, the dual-bath configuration simultaneously accelerates convergence, reduces the total gate cost, and improves robustness to hardware noise.

\section*{Acknowledgment}
The authors thank Giovanna Morigi, Sayan Roy, Emma C. King, Vinayak Pendse and Francesco Mattiotti for their insightful remarks and are grateful for the stimulating discussions. We gratefully acknowledge financial support from the Quantum Initiative Rhineland-Palatinate QUIP, the Research Initiative Quantum Computing for AI (QC-AI), and the Center for Quantum Technologies (QuTe). We also acknowledge with gratitude the support of IQM Quantum Computers in providing the Resonance credits used to perform the hardware experiments in this work as part of the IQM Research Award 2026. 

AI tools, including Claude Opus 5.5 and OpenAI ChatGPT 6 Astra, were used for spelling and grammar checks, text shortening, and stylistic revisions to improve the readability of the text. Claude Opus 5.5 also assisted in creating the TikZ circuit diagrams shown in Fig.~(\ref{fig:full_circuit}) and Fig.~(\ref{fig:lightcone}). All authors carefully reviewed the resulting text and figures, verified their accuracy, and take full responsibility for the final manuscript.

\appendix
\section{Quasiparticle picture of TFIM}
\label{app:appendix_a}

To investigate how the location of ancillary baths affects dissipative ground-state preparation, we consider the transverse-field Ising model defined in Eq.~\eqref{Eq:TFIM_Hamiltonian}, with local coupling operators $A_j=Z_j$. Using the Jordan--Wigner transformation, we analyze the structure of the corresponding filtered jump operators defined in Eq.~\eqref{eq:jump}, distinguishing between boundary and bulk couplings.

\subsection{Fermionic representation}

We first apply a global Hadamard transformation,
$U_{\mathrm H}=\mathrm{Had}^{\otimes N}$, under which
$Z_j\mapsto X_j$ and $X_j\mapsto Z_j$.
The transformed Hamiltonian and coupling operators become
\begin{align}
    H' &= U_{\mathrm H} H U_{\mathrm H}^{\dagger}
    = J\sum_{j=0}^{N-2}X_jX_{j+1}
      +h_x\sum_{j=0}^{N-1}Z_j,
    \label{eq:app_rotated_H}\\
    A'_j &= U_{\mathrm H}A_jU_{\mathrm H}^{\dagger}
    =X_j.
    \label{eq:app_rotated_A}
\end{align}
Since this is a unitary transformation, it leaves the
spectrum of the Hamiltonian and the singular values of
the jump operators unchanged.

Following the Jordan--Wigner transformation, we define
the fermionic operators
\begin{equation}
    c_j =
    \left(\prod_{k=0}^{j-1}Z_k\right)
    \frac{X_j-iY_j}{2},
    \qquad
    c_j^\dagger =
    \left(\prod_{k=0}^{j-1}Z_k\right)
    \frac{X_j+iY_j}{2}.
    \label{eq:app_jw}
\end{equation}
The associated Majorana operators are
\begin{align}
    \gamma_j &=
    \frac{c_j+c_j^\dagger}{\sqrt{2}}
    =
    \frac{1}{\sqrt{2}}
    \left(\prod_{k=0}^{j-1}Z_k\right)X_j,
    \\
    \gamma_{j+N} &=
    \frac{i(c_j-c_j^\dagger)}{\sqrt{2}}
    =
    \frac{1}{\sqrt{2}}
    \left(\prod_{k=0}^{j-1}Z_k\right)Y_j.
\end{align}

Under this transformation, the Hamiltonian takes a
quadratic form in the $2N$ Majorana operators,
\begin{equation}
    H' =
    -2iJ\sum_{j=0}^{N-2}\gamma_{j+N}\gamma_{j+1}
    -2ih_x\sum_{j=0}^{N-1}\gamma_j\gamma_{j+N}.
    \label{eq:app_majorana_H}
\end{equation}
An important property of quadratic Hamiltonians is
that the Heisenberg evolution of a Majorana operator
remains linear in Majorana operators,
\begin{equation}
    e^{iH's}\gamma_p e^{-iH's}
    =\sum_{q=0}^{2N-1}R_{pq}(s)\gamma_q,
    \label{eq:app_majorana_evolution}
\end{equation}
where $R(s)$ is a real orthogonal matrix describing
the evolution in Majorana space. This property underlies the quasi-free description of boundary-driven dissipative dynamics discussed
in Ref.~\cite{zhan_rapid_2026}.

\subsection{Coupling at the boundaries}

We first consider a bath coupled to the left boundary at site $j=0$. From the Jordan--Wigner transformation,
\begin{equation}
    A'_0=X_0=\sqrt{2}\,\gamma_0.
\end{equation}
Since the Hamiltonian is quadratic, the filtered jump
operator remains linear in Majorana operators,
\begin{equation}
    K'_0=\sum_{q=0}^{2N-1}\zeta_{0q}\gamma_q,
\end{equation}
where $\zeta_{0q}$ are the coefficients determined
by the Hamiltonian and the filter function.
The quadratic Hamiltonian can be diagonalized through
a Bogoliubov transformation~\cite{bogoliubov1947theory},
\begin{equation}
    H'=\sum_{k=1}^{N}\varepsilon_k
    \left(b_k^\dagger b_k-\frac{1}{2}\right),
\end{equation}
where $\varepsilon_k>0$ are the quasiparticle
excitation energies and $b_k$ are the corresponding
fermionic annihilation operators, linear in Majorana operators.
Expressing the boundary coupling operator in this
basis and applying the frequency filter yields
\begin{equation}\label{eq:A11}
    K'_0=\sum_{k=1}^{N}
    \left[
    u_kF(-\varepsilon_k)b_k
    +u_k^*F(\varepsilon_k)b_k^\dagger
    \right].
\end{equation}
The coefficients $u_k$ are those of the expansion
$A'_0=\sum_{k=1}^{N}(u_kb_k+u_k^*b_k^\dagger)$.
Since $(A'_0)^2=\mathbb{1}$, they satisfy the sum
rule $\sum_{k=1}^{N}|u_k|^2=1$. For an ideal downward-only filter, $F(\varepsilon_k)=0$ for all $k$, and hence the creation terms in Eq.~\eqref{eq:A11} vanish. The resulting jump operator can be written as~\cite{zhan_rapid_2026}
\begin{equation}
    K'_0=\sigma c,
    \qquad
    \sigma^2=\sum_{k=1}^{N}|u_k|^2|F(-\varepsilon_k)|^2,
    \label{eq:app_K0_single_mode}
\end{equation}
where $c=\sigma^{-1}\sum_{k=1}^{N}u_kF(-\varepsilon_k)b_k$
is a canonical fermionic annihilation operator, assuming
$\sigma>0$. Thus, each jump reduces the total Bogoliubov
quasiparticle number by one. The single-mode structure also determines the singular-value
spectrum of $K'_0$. Since $c^\dagger c$ is a projection,
$(K'_0)^\dagger K'_0=\sigma^2c^\dagger c$ has eigenvalues
$0$ and $\sigma^2$, yielding a single distinct non-zero
singular value $\sigma$.

At the opposite boundary, $j=N-1$, the coupling
operator can be expressed as
\begin{equation}
    A'_{N-1}=i\sqrt{2}\,P\gamma_{2N-1},
    \qquad
    P=\prod_{j=0}^{N-1}Z_j,
\end{equation}
where $P$ is the global fermionic parity operator.
Since $[P,H']=0$, the parity operator can be taken
outside the filtered evolution, giving
\begin{equation}
    K'_{N-1}=P\sum_{q=0}^{2N-1}\zeta_{Rq}\gamma_q,
\end{equation}
where $\zeta_{Rq}$ are the filter-dependent
coefficients for the right-boundary coupling.
Thus, the right-boundary jump operator is linear
in Majorana operators up to a global parity factor.
As for the left boundary, an ideal downward-only
filter retains only quasiparticle annihilation terms.
The right-boundary jump operator therefore also
removes a single Bogoliubov quasiparticle, with
the parity factor affecting the phase of
the resulting state.

Since $P$ is unitary, it does not affect the
singular values of the jump operator. Moreover,
reflection symmetry of the open-chain Hamiltonian
ensures that the jump operators at the two
boundaries have identical singular-value spectra
when the same filter is used.

This singular-value structure has direct implications
for the choice of the dissipative timestep discussed
in Sec.~III~B. Under an ideal downward-only filter,
all non-zero singular values of the boundary jump
operators are equal to $\sigma$, with multiplicity
$2^{N-1}$. Consequently, the choice
$\tau_{\max}=(\pi/2\sigma)^2$ in
Eq.~\eqref{Eq:max_dissipation_tau} simultaneously
maximizes the jump probability for all non-zero
singular modes.

By the normalization condition on the coefficients
$u_k$, $\sigma^2$ in Eq.~\eqref{eq:app_K0_single_mode}
is a weighted average of $|F(-\varepsilon_k)|^2$
over the quasiparticle excitation energies.
For the TFIM parameters considered in this work,
these energies lie within the single-particle band
\begin{equation}
    2\big||h_x|-|J|\big|
    \leq \varepsilon_k
    \leq 2(|h_x|+|J|),
\end{equation}
whose bounds are independent of $N$. Consequently,
\begin{equation}
    \min_{\varepsilon\in\mathcal{B}}|F(-\varepsilon)|^2
    \leq \sigma^2
    \leq
    \max_{\varepsilon\in\mathcal{B}}|F(-\varepsilon)|^2,
\end{equation}
where $\mathcal{B}$ denotes the single-particle energy
band. Provided that the filter response is bounded
away from zero throughout this band, both $\sigma$
and $\tau_{\max}$ admit system-size-independent bounds.

For the discretized filter used in this work
($S_s=2$, $M_s=5$), neglecting residual quasiparticle
creation terms, we obtain $\tau_{\max}=2.971$,
$2.975$, and $2.978$ for $N=4$, $5$, and $8$,
respectively, approaching $2.979$ for $N\geq16$.
For $N=5$, the full singular-value decomposition
of the discretized jump operator yields
$\tau_{\max}=2.970$, with the relative discrepancy
of less than $0.2\%$ arising from the residual
creation terms neglected in
Eq.~\eqref{eq:app_K0_single_mode}.
These results indicate that, for the parameters
considered here, the boundary-coupling timestep
exhibits only a weak dependence on the system size
and can be reliably estimated from a $2N\times2N$
free-fermion calculation or from a sufficiently
small system.

\subsection{Coupling to bulk sites}

We now consider a coupling operator at an interior
site, $0<j<N-1$. The Jordan--Wigner transformation
gives
\begin{equation}
    A'_j=X_j
    =\sqrt{2}
    \left(\prod_{k=0}^{j-1}Z_k\right)\gamma_j.
\end{equation}
Substituting $Z_k=-2i\gamma_k\gamma_{k+N}$, we obtain
\begin{equation}
    A'_j=
    \sqrt{2}(-2i)^j
    \left(\prod_{k=0}^{j-1}\gamma_k\gamma_{k+N}\right)\gamma_j.
    \label{eq:app_bulk_coupling}
\end{equation}
Unlike the first-boundary operator, an interior
coupling is therefore represented by a product
of $2j+1$ Majorana operators in the chosen
Jordan--Wigner ordering. Under the quadratic Hamiltonian, individual Majorana operators evolve linearly, whereas products of Majorana
operators generally evolve into linear combinations of
products of the same order. Consequently, the filtered
jump operator for a bulk coupling generally
contains higher-order fermionic terms and cannot be
expressed as a linear combination of Majorana operators.

In the Bogoliubov basis, these higher-order
operators generate multi-quasiparticle transitions
involving products of creation and annihilation
operators. The resulting dissipative dynamics
therefore does not, in general, admit the same
quasi-free reduction as boundary coupling. Importantly, the presence of higher-order fermionic terms does not itself imply slower
cooling. Such terms can introduce additional
relaxation pathways. Their effect on convergence
depends on the transition matrix elements,
the filter function, and the resulting
dissipative generator. In particular, a finite discretization of the
filter may produce residual positive-frequency
components, allowing unwanted upward transitions.
The relative importance of these contributions
depends on the coupling site and can affect
the attainable energy in a finite-step
implementation.

\section{Number and placement of ancillary baths}
\label{app:appendix_b}

We investigate how the number and placement of ancillary baths affect dissipative ground-state preparation for the $N=5$ transverse-field Ising model. Building on the fermionic analysis in Appendix~\ref{app:appendix_a}, we compare configurations with two and three ancillas coupled to different sites of the chain. Unless stated otherwise, we use the generalized-normal filter with parameters $(\beta,\mu,\alpha)=(6,-5,5)$, truncation parameter $S_s=2$, Hamiltonian evolution timestep $\Delta_t=0.25$, and $10^4$ measurement shots per basis. Each ancilla couples through a local operator $A_a=Z_{j_a}$, with the dissipative timestep $\tau_{j_a}$ determined by Eq.~\eqref{Eq:max_dissipation_tau}. For ancillas coupled to the same site, we use $\tau_{\max}/2$ for each bath. We compare a coarse temporal grid with $M_s=5$ ($\Delta_s=0.4$) and a finer grid with $M_s=15$ ($\Delta_s\approx0.13$). Plateau energies are estimated by averaging over dissipative layers $m=16$ to $20$. Table~\ref{tab:dual_ancilla_table} summarizes the plateau energies and the number of dissipative layers required to reach 85\% of $E_0$ for the different ancilla configurations on the coarse grid.

\subsection{Two ancillas}

\begin{figure}
    \centering
    \includegraphics[width=\columnwidth]{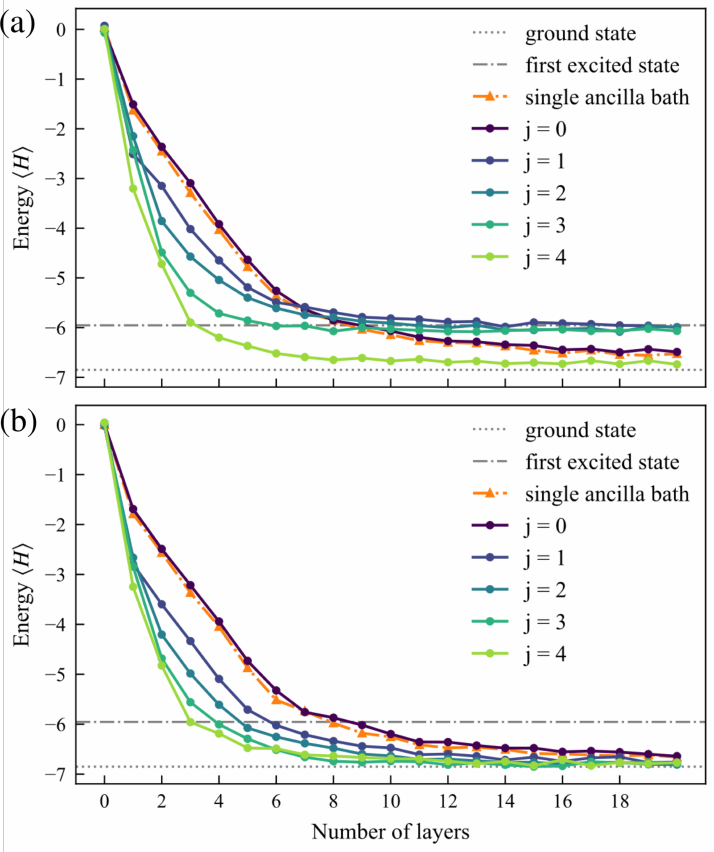}
    \caption{Performance protocol for dual ancilla baths with one ancilla fixed at $j=0$ and other for varying $j$. The top plot corresponds to coarser grid with $M_s=5$ and the bottom plot for finer grid $M_s = 15$. The orange dashdot curve corresponds to protocol with single ancilla bath at $j=0$. All the curves are obtained via Aer simulator.}
    \label{fig:2ancilla_diffloc}
\end{figure}

We fix the first ancilla at the left end ($j=0$) and move the second over $j=0,\dots,4$ (Fig.~\ref{fig:2ancilla_diffloc}).

\paragraph*{Energy plateaus}
For the coarse grid, the configuration with ancillas at opposite boundaries ($j=4$) reaches 97.9\% of $E_0$, whereas configurations with the second ancilla coupled to a bulk site ($j=1,2,3$) saturate at approximately 87--88\%. This difference is consistent with aliased heating arising from the finite discretization of the filter, as discussed in Sec.~\ref{sec:filter}. A bulk operator $Z_j$ drives multi-quasiparticle transitions up to $\omega\approx9$. These lie above the Nyquist frequency $\pi/\Delta_s=7.85$ and fall inside the aliased copy of the cooling band. The ground state is then no longer dark. On the fine grid the aliased copy lies beyond the many-body bandwidth. All configurations with distinct sites then reach 98--99\% of $E_0$ within 20 layers, with no systematic difference between bulk and end placements. The plateau on the coarse grid is therefore a discretization artifact, not an intrinsic disadvantage of bulk coupling.

\paragraph*{Convergence speed}
Placing both ancillas at the same end ($j=0$) reproduces the single-ancilla protocol i.e., both reach 85\% of $E_0$ after 8 layers and plateau at 95\%. Both ancillas then couple to the same fermionic mode, so the second adds no new cooling channel. Placing the second ancilla at the opposite end gives the fastest initial cooling i.e., 85\% of $E_0$ after 3 layers, compared with 5--11 layers for bulk placements (see Table.~\ref{tab:dual_ancilla_table}). The two boundary jump operators act on nearly orthogonal fermionic modes, each therefore removes energy independently, and the energy extracted per layer doubles compared with a single ancilla. A bulk ancilla partially overlaps with the mode already cooled at the left end, and its singular values are spread, so it adds less per layer.

\subsection{Three ancillas}

\begin{figure}
    \centering
    \includegraphics[width=\columnwidth]{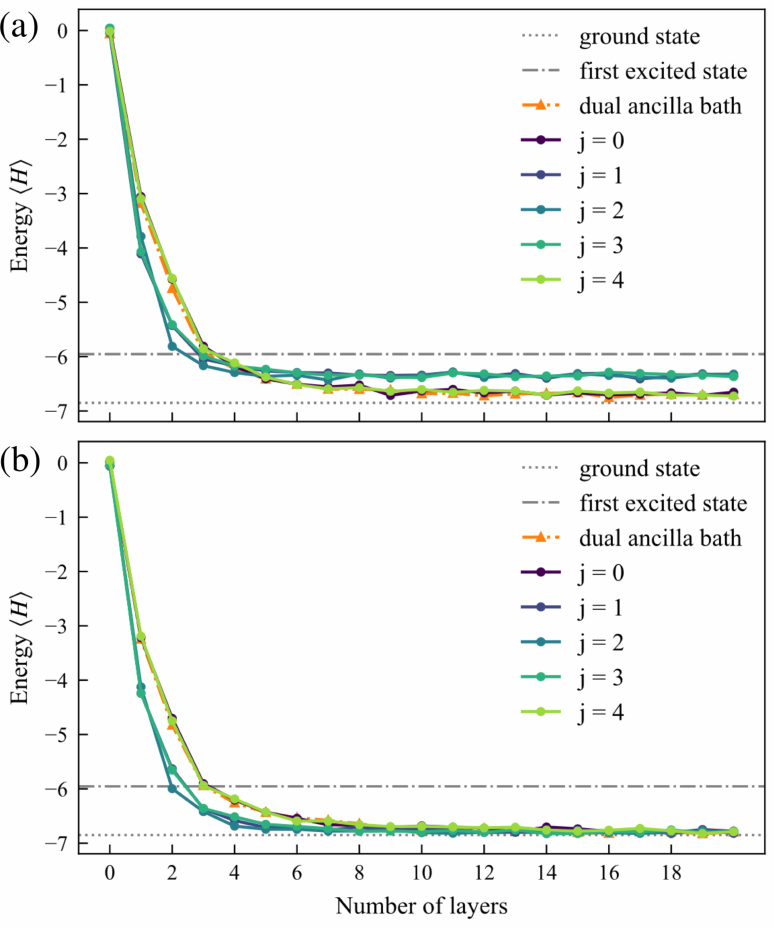}
    \caption{Performance of protocol for three ancilla baths with two ancillas fixed at $j=0, N-1$ and the third ancilla for varying $j$. The top plot corresponds to coarser grid with $M_s=5$ and the bottom plot for finer grid $M_s = 15$. The orange dashdot curve corresponds to protocol with dual ancilla bath at $j=0, N-1$. All the curves are obtained via Aer simulator.}
    \label{fig:3ancilla_diffloc}
\end{figure}

We fix two ancillas at both ends and move a third over $j=0,\dots,4$ (Fig.~\ref{fig:3ancilla_diffloc}). A third ancilla at $j=0$ or $j=4$ couples to a mode that is already cooled. With $\tau_{\max}/2$ on the doubled end, the configuration is equivalent to the two-ancilla end-to-end protocol. Correspondingly, it shows almost no improvement. In contrast, placing the third ancilla at a bulk site accelerates initial cooling, reaching 85\% of $E_0$ after three layers and achieving lower energies after the first layer than the two-ancilla configuration. The lower plateau energies on the coarse grid are consistent with aliased heating, as discussed above. Increasing the grid resolution suppresses this effect, allowing the three-ancilla configurations with bulk coupling to reach approximately 99\% of $E_0$.

\subsection{Summary}

For the five-qubit TFIM, two ancillas at opposite ends of the chain are the optimal configuration. Adding an ancilla at an already coupled boundary provides little additional benefit, whereas introducing a bulk ancilla can accelerate initial cooling. On the coarse grids feasible with current quantum hardware, however, bulk couplings are susceptible to aliasing-induced heating, resulting in higher energy plateaus. Refining the temporal grid improves convergence toward the ground state but increases circuit depth. Moreover, bulk coupling reduces opportunities for light-cone cancellation, further increasing the circuit cost. These results highlight the trade-off between enhanced cooling and implementation complexity when choosing the number and placement of ancillary baths

\begin{table}[t]
\centering
\caption{Multi ancilla protocol for $N=5$ and $M_s=5$. Plateau energies are given in \% of $E_0$, averaged over layers 16--20, together with the first layer at which 85\% of $E_0$ is reached. The datapoints are obtained from summarizing Fig.~\ref{fig:2ancilla_diffloc} and Fig.~\ref{fig:3ancilla_diffloc}.}
\label{tab:dual_ancilla_table}
\setlength{\tabcolsep}{5pt}
\begin{tabular}{c cc cc}
\hline
 & \multicolumn{2}{c}{two ancillas} & \multicolumn{2}{c}{three ancillas}\\
$j$ & plateau & 85\% at $m$& plateau & 85\% at $m$\\
\hline
0 (same site) & 94.3 & 8 & 97.6 & 4\\
1 & 86.9 & 11 & 92.7 & 3\\
2 & 87.9 & 9  & 92.7 & 3\\
3 & 88.3 & 5  & 92.4 & 3\\
4 (opposite end) & \textbf{97.9} & \textbf{3} & 97.7 & 3\\
\hline
\end{tabular}
\end{table}

\section{Gate count of the single- and dual-bath protocols}
\label{app:appendix_c}

Although the dual-bath protocol accelerates convergence toward the ground state, introducing a second ancilla increases the gate count per dissipative layer. Here, we derive analytical expressions for the two-qubit gate counts of the single- and dual-bath protocols after light-cone cancellation and determine the conditions under which faster convergence compensates for the additional gates per layer. Each $R_{ZZ}$ gate is decomposed into two CZ gates in the native gate set of IQM hardware. Throughout this analysis, we consider $N_t=4$ (see Fig.~\ref{fig:full_circuit}\,a), corresponding to two Trotter steps for each of the intermediate Hamiltonian evolutions. For an open chain of $N$ qubits, we index the $N-1$ nearest-neighbor couplings by $k=0,\ldots,N-2$, starting from the left boundary.

\begin{figure}[!h]
\centering
\includegraphics{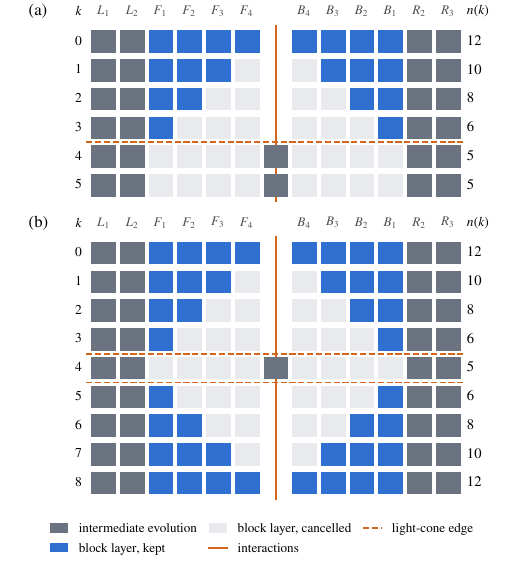}
\caption{Merged $R_{ZZ}$ layers on each coupling $k$ for $M_s=2$. Columns are the layers in circuit order: intermediate evolution ($L$, $R$), forward sweep ($F_p$), interactions (vertical line), and backward sweep ($B_p$). Mirror pairs $F_p$, $B_p$ survive only within the light cone of the bath (blue) and cancel outside it (light grey). Beyond the light-cone edge at $k=2M_s$, only the five layers of the intermediate evolution remain. The last column gives $n(k)$ of Eq.~\eqref{eq:app_nk}. (a)~Single bath at qubit $0$, $N=7$. (b)~Baths at both ends, $N=10$: each bath opens its own light cone, and each bond is counted from the nearer bath.}
\label{fig:app_lightcone}
\end{figure}

\subsection{Gate count before light-cone cancellation}

As discussed in Sec.~\ref{Sec:methods} A, the dissipative block $\mathcal{W}(\sqrt{\tau})$ contains $4M_s+1$ ancilla--system interactions per bath, each implemented by a single $R_{ZZ}$ gate according to Eq.~(\ref{Eq:U_l}).

The dissipative block also contains $4M_s$ Hamiltonian evolutions $e^{\pm iH\Delta_s}$, each comprising two $R_{ZZ}$ half-layers, as shown in Fig.~\ref{fig:full_circuit}(d). Since the ancilla--system interactions between successive Hamiltonian evolutions act on the system through $Z$ operators, they commute with the $R_{ZZ}$ gates, allowing adjacent half-layers to merge. This results in $4M_s$ merged $R_{ZZ}$ layers per ancilla.

The intermediate Hamiltonian evolutions $e^{-iH\Delta_t}$, with $N_t=4$, comprise two Trotter steps on either side of the dissipative block. Each Trotter step contains two $R_{ZZ}$ half-layers, which merge into three $R_{ZZ}$ layers per side. The innermost layer on each side further merges with the adjacent layer of the dissipative block. Consequently, before light-cone cancellation, each nearest-neighbor coupling is associated with a total of $(4M_s+4)$ $R_{ZZ}$ layers.

\subsection{Gate count after light-cone cancellation}

We first consider a single ancillary bath coupled to the left boundary of the chain. The forward and backward Hamiltonian evolutions within the dissipative block $\mathcal{W}(\sqrt{\tau})$ exhibit a mirrored structure, with each $R_{ZZ}$ layer $F_p$ in the forward sweep having a corresponding inverse layer $B_p=F_p^\dagger$ in the backward sweep, as illustrated in Fig.~\ref{fig:app_lightcone}. Light-cone cancellation (LCC) eliminates pairs of inverse gates outside the region influenced by the ancilla--system interactions.

The innermost pair survives only on the first nearest-neighbor coupling ($k=0$) due to the interaction of the ancilla with the first system qubit. With each successive pair moving outward from the center of the dissipative block, the region in which cancellation is prevented extends by one nearest-neighbor coupling. Thus, within this region, the $k$ innermost pairs cancel on coupling $k$, reducing the number of $R_{ZZ}$ layers by $2k$.

After the merging and cancellation of the $R_{ZZ}$ half-layers described above, the forward and backward sweeps contain $2M_s$ pairs of inverse $R_{ZZ}$ layers. Consequently, only the first $2M_s$ nearest-neighbor couplings retain contributions from the dissipative block. For $k\geq 2M_s$, the contributions from the dissipative block cancel completely on the corresponding coupling, allowing the adjacent $R_{ZZ}$ layers of the intermediate Hamiltonian evolutions to merge. Since each intermediate evolution contributes three $R_{ZZ}$ layers for $N_t=4$, this merging leaves a total of five layers. The number of remaining $R_{ZZ}$ gates on coupling $k$ is therefore
\begin{equation}
n(k)=\max\left(4M_s+4-2k,\;5\right).
\label{eq:app_nk}
\end{equation}
We refer to the spatial region comprising of first $2M_s$ nearest-neighbor couplings as the \emph{light cone} of the bath. The total number of $R_{ZZ}$ gates per dissipative layer for the single-bath protocol is obtained by summing the contributions from the ancilla--system interactions and the remaining gates on all nearest-neighbor couplings:
\begin{equation}
  \mathrm{RZZ}_1=(4M_s+1)+\sum_{k=0}^{N-2}n(k).
\end{equation}
If the entire chain lies within the light cone, i.e., $N-1\leq 2M_s$, this expression reduces to
\begin{equation}
  \mathrm{RZZ}_1=(4M_s+1)+(N-1)(4M_s+4)-(N-1)(N-2),
\end{equation}
where the last term accounts for the reduction in gate count due to light-cone cancellation, given by $\sum_{k=0}^{N-2}2k$. For longer chains, each additional nearest-neighbor coupling outside the light cone contributes only five $R_{ZZ}$ gates from the intermediate Hamiltonian evolutions. Consequently, the total gate count increases linearly with system size beyond the light-cone extent.

Introducing a second ancillary bath at the opposite boundary of the chain doubles the number of ancilla--system interactions and introduces a second light cone, as illustrated in Fig.~\ref{fig:app_lightcone}(b). The number of remaining $R_{ZZ}$ gates on each nearest-neighbor coupling is determined by its distance $d_k=\min(k,N-2-k)$ from the nearer bath. The total number of $R_{ZZ}$ gates per dissipative layer for the dual-bath protocol is therefore
\begin{equation}
\mathrm{RZZ}_2=2(4M_s+1)+\sum_{k=0}^{N-2}n(d_k).
\end{equation}

The presence of the second light cone reduces the number of gates that can be eliminated through light-cone cancellation. Together, the two light cones extend over at most $4M_s$ nearest-neighbor couplings. For $N-1>4M_s$, the light cones no longer overlap, and the additional gate count associated with the second bath becomes independent of the system size:
\begin{equation}
\mathrm{RZZ}_2-\mathrm{RZZ}_1=(4M_s+1)+4M_s^2,
\end{equation}
where the first term accounts for the additional ancilla--system interactions and the second for the Hamiltonian evolution gates retained within the second light cone.

\subsection{Gate-count overhead and convergence speed-up}

Table~\ref{tab:app_gatecount} compares the analytical gate counts derived above with the transpiled gate counts of Fig.~\ref{fig:2anc_1anc_gate_counts_scaling} for $M_s=5$. The relative gate-count overhead per dissipative layer introduced by the second bath is quantified by
\begin{equation}
  r(N)=\frac{\mathrm{CZ}_2}{\mathrm{CZ}_1},
\end{equation}
where $\mathrm{CZ}_i=2\,\mathrm{RZZ}_i$ denotes the analytical CZ gate count for the protocol with $i$ baths.

\begin{table}[t]
\centering
\caption{Analytical and transpiled CZ gate counts per dissipative layer for the single- and dual-bath protocols with $M_s=5$ and $N_t=4$. The transpiled gate counts are obtained from Fig.~\ref{fig:2anc_1anc_gate_counts_scaling}. Transpilation into the IQM native gate set eliminates an additional four and six CZ gates for the single- and dual-bath protocols, respectively. The ratio $r(N)$ is calculated from the analytical gate counts.}
\label{tab:app_gatecount}
\begin{tabular}{c cc cc c}
\hline
 & \multicolumn{2}{c}{single bath} & \multicolumn{2}{c}{dual bath} & \\
$N$ & formula & transpiled & formula & transpiled & $r(N)$\\
\hline
5  & 210 & 206 & 268 & 262 & 1.27\\
10 & 330 & 326 & 452 & 446 & 1.37\\
14 & 372 & 368 & 564 & 558 & 1.52\\
\hline
\end{tabular}
\end{table}

Although the dual-bath protocol requires more gates per dissipative layer, its faster convergence can reduce the total number of gates required to reach a given target energy. Let $m_1$ and $m_2$ denote the number of dissipative layers required by the single- and dual-bath protocols, respectively, to reach the same target energy. Defining the convergence speed-up as $s=m_1/m_2$, the ratio of the total CZ gate counts is given by
\begin{equation}
  \frac{m_2\,\mathrm{CZ}_2}{m_1\,\mathrm{CZ}_1}
  =\frac{r(N)}{s}.
\end{equation}
The dual-bath protocol therefore requires fewer CZ gates in total whenever $s>s^*$, where $s^*=r(N)$ defines the break-even speed-up.

For $M_s=5$, this threshold $s^*$ increases from approximately $1.27$ at $N=5$ to a maximum of approximately $1.56$ at $N=19$. Once the light cones become spatially separated, the additional gate count introduced by the second bath becomes independent of $N$. Consequently, the relative gate-count overhead decreases, and $s^*$ approaches unity with increasing system size. Within the considered gate-count model, a convergence speed-up exceeding approximately $1.56$ is therefore sufficient for the dual-bath protocol to reduce the total CZ gate count across all system sizes.

For $N=5$, reaching $85\%$ of $E_0$ requires eight dissipative layers with a single bath and three layers with two baths, corresponding to a convergence speed-up of approximately $2.7$. Using the transpiled gate counts in Table~\ref{tab:app_gatecount}, the total number of CZ gates decreases from $8\times206=1648$ to $3\times262=786$, corresponding to a reduction of approximately $52\%$. This accelerated convergence is consistent with the fermionic picture developed in Appendix~\ref{app:appendix_a}, where the two boundary jump operators provide complementary quasiparticle relaxation channels.

These results demonstrate that the accelerated cooling achieved by introducing a second boundary-coupled bath can compensate for the additional gates required per dissipative layer, substantially reducing the total gate count required to reach a given target energy.

\bibliographystyle{apsrev4-2}
\bibliography{ref}

\end{document}